\documentclass[10pt,aps,showpacs,nofootinbib,prd,aps,epsf,floats,
               amsmath,amssymb,amsfonts,axodraw,floatfix,graphicx,twocolumn]{revtex4-1}
\usepackage{amsmath, amssymb}
\usepackage{multirow}
\usepackage{paralist}
\newcommand{\mathsym}[1]{{}}

\usepackage{graphicx}
\usepackage{amsmath}
\usepackage{amssymb}
\usepackage{amsmath}
\usepackage{multirow}
\usepackage{paralist}
\usepackage{slashed}
\usepackage{amsfonts}
\usepackage{hyperref} 
\usepackage{graphicx}
\usepackage{amsmath}
\usepackage{amssymb}
\usepackage{mathrsfs}
\newsavebox{\PSLASH}
 \sbox{\PSLASH}{$p$\hspace{-1.8mm}/}
 
\renewcommand{\theequation}{\thesection.\arabic{equation}}
\newcounter{saveeqn}
\newcommand{\add}{\addtocounter{equation}{1}}
\newcommand{\alphaeqn}{\setcounter{saveeqn}{\value{equation}}%
\setcounter{equation}{0}%
\renewcommand{\theequation}{\mbox{\thesection.\arabic{saveeqn}{\alpha{equation}}}}}
\newcommand{\reseteqn}{\setcounter{equation}{\value{saveeqn}}%
\renewcommand{\theequation}{\thesection.\arabic{equation}}}

 \newsavebox{\notrightarrow}
 \sbox{\notrightarrow}{$\to$\hspace{-4mm}/}
 
 \newsavebox{\PARTIALSLASH}
 \sbox{\PARTIALSLASH}{$\partial$\hspace{-1.6mm}/}
 
 \newsavebox{\ASLASH}
 \sbox{\ASLASH}{$A$\hspace{-2.1mm}/}
 
 \newsavebox{\KSLASH}
 \sbox{\KSLASH}{$k$\hspace{-1.8mm}/}
 
 \newsavebox{\LSLASH}
 \sbox{\LSLASH}{$\ell$\hspace{-1.8mm}/}
 
 \newsavebox{\QSLASH}
 \sbox{\QSLASH}{$q$\hspace{-1.8mm}/}
 
 \newsavebox{\DSLASH}
 \sbox{\DSLASH}{$D$\hspace{-2.2mm}/}
 
 \newsavebox{\DbfSLASH}
 \sbox{\DbfSLASH}{${\mathbf D}$\hspace{-2.8mm}/}
 
 \newsavebox{\DELVECRIGHT}
 \sbox{\DELVECRIGHT}{$\stackrel{\rightarrow}{\partial}$}
 
 \newcommand{\blue}{\IfColor{\textCadetBlue}{}}
\newcommand{\black}{\IfColor{\textBlack}{}}
\newcommand{\red}{\IfColor{\textRed}{}}
\newcommand{\green}{\IfColor{\textOliveGreen}{}}
\newcommand{\lil}{\IfColor{\textRedViolet}{}}

\newcommand{\bs}{\boldsymbol}

\makeatother
\usepackage[T1]{fontenc}
\usepackage[latin9]{inputenc}
\usepackage{amsmath}
\usepackage{amssymb}
\usepackage{graphicx}
\usepackage{dcolumn}
\usepackage{verbatim}

\usepackage{orcidlink}

\begin{document}
\title{Chiral vortical conductivities and the moment of inertia of a rigidly rotating Fermi gas}
\author{M. Abedlou Ahadi~~}\email{mohammad_abedlouahadi@physics.sharif.edu}
\author{N. Sadooghi\,~\orcidlink{0000-0001-5031-9675}~~}\email{Corresponding author: sadooghi@physics.sharif.ir}
\affiliation{Department of Physics, Sharif University of Technology,
P.O. Box 11155-9161, Tehran, Iran}
\begin{abstract}
We determine the chiral vortical conductivities, as well as the orbital and spin moment of inertia of a charged,
chiral, and rigidly rotating free Fermi gas. To this purpose, we begin by calculating the vacuum
expectation values of a vector and axial vector current using the free fermion propagator
in this medium. This propagator is derived by employing the Fock-Schwinger method based on the solutions of the Dirac equation in the presence of rotation and finite axial
chemical potential. We present a complete derivation of these solutions. We demonstrate that in the first approximation, the chiral vortical
conductivity associated with the vector current is proportional to the product of the vector and
axial chemical potentials. In contrast, the chiral vortical conductivity related to
the axial vector current depends on the temperature, the vector, and the axial vector
chemical potential squares. We use the relation between the axial vector current and the angular momentum density associated with spin to determine the spin and the orbital moment of inertia of a rigidly rotating Fermi gas in a charged and chirally imbalanced medium separately.
In addition, we compute the total moment of inertia by utilizing the methods presented in the first part of the paper and show that the orbital moment of inertia of a free fermion gas vanishes.
\end{abstract}

\maketitle
\section{Introduction}\label{sec1}
\setcounter{equation}{0}
As is well known, the interplay between the chiral anomaly and a uniform magnetic field and/or rotation (vorticity) leads to various important transport phenomena such as chiral magnetic effect (CME) \cite{warringa2008, fukushima2008} and chiral vortical effect (CVE) \cite{zhitnisky}. These phenomena principally occur in media that exhibits chirality imbalance, and are characterized by the formation of vector or axial vector currents that are aligned with the direction of an external magnetic field $\boldsymbol{B}$, or angular velocity $\boldsymbol{\Omega}$. In recent years, extensive theoretical efforts have been made to understand various aspects of these effects. Additionally, modern heavy ion collision experiments conducted at the Relativistic Heavy Ion Collider (RHIC) and the Large Hadron Collider (LHC) provide a unique opportunity to investigate the experimental consequences of these phenomena (see \cite{kharzeev2015, shovkovy2015, kharzeev2024,voloshin2025} for recent reviews). The reason is that extremely large magnetic field up to $10^{18}-10^{20}$ Gau\ss~ \cite{skokov} and vorticity with the angular velocity up to $10^{22}$ Hz \cite{voloshin2017} can be created in the early stages of these collisions (see also recent papers \cite{is2025} and \cite{star2023,kharzeev2025}). All these efforts aim to investigate the nontrivial topological sector of quantum chromodynamics. Theoretical studies primarily focus on determining the anomalous transport coefficients, which act as proportionality factors between vector or axial vector currents and $\boldsymbol{B}$ or $\boldsymbol{\Omega}$. Because of the specific properties of these currents, as well as the behavior of $\boldsymbol{B}$ and $\boldsymbol{\Omega}$ under parity and time reversal transformation, the transport coefficients associated with CME and CVE are odd under parity. The corresponding currents are thus nondissipative \cite{kharzeev,kharzeev2015}.
Exploring experimental consequences of CME and CVE also builds an active research direction in condensed matter physics \cite{cve-cond}, astrophysics and cosmology \cite{cve-astro}.
\par
Modern literature offers various approaches to determine nondissipative anomaly induced transport coefficients. These methods include relativistic (magneto)hydrodynamics \cite{son, sadooghi2016}, gauge/gravity correspondence \cite{landsteiner2011,landsteiner2013}, and chiral kinetic theory \cite{liao}. In \cite{ambrus2014}, different condensates, including vector and axial vector currents, are computed using a technique originally introduced in \cite{vilenkin1980}. In this method, the quantization relation of fermions is inserted in the definition of the condensates. By applying the Kubo-Martin-Schwinger formalism, the Fermi-Dirac distribution function appears in the calculations. The final results coincide with those obtained through other methods. Another technique for determining fermionic condensates involves the free fermion two-point function. This method performs a limit to make the propagator local and carries out a trace over Dirac matrices to convert it into a scalar quantity. Using, in particular, the free fermion propagator at finite temperature, the resulting expression yields a condensate that is consistent with results from other methods. In \cite{sadooghi2014}, the same approach is used to calculate the chiral condensate in the presence of a constant and an exponentially decaying magnetic field. In the present paper, we use, for the first time this method to determine the chiral vortical conductivities and the moment of inertia of a charged, chiral and rigidly rotating free Fermi gas.
\par
In the first part of this paper, we focus on the CVE and determine chiral vortical conductivities corresponding to vector and axial vector currents in a medium with a net charge and chirality. To this purpose, we start with a Dirac Lagrangian in a medium with finite vector (electric) and axial chemical potential. In addition, we assume that the free fermion gas described by this Lagrangian is under a rigid rotation. The rotation is introduced by a metric originally used in \cite{rot1} to describe a rigid rotation. Many other works studying the effect of rotation on fermionic and bosonic systems are based on this metric \cite{fukushima2015,rot3,rot4,rot5,rot6,rot7,huang,rot8,rot9,ayala2021,rot10,siri2024-1,siri2024-2,siri2024-3,rot11,rot12,hua2024}. We present a detailed derivation of the solution of the Dirac equation arising from this Lagrangian. Our solution agrees, up to a normalization factor, with the result recently presented in \cite{hua2024}, where no details are provided. By utilizing the Fock-Schwinger method, introduced in \cite{iablokov2019,iablokov2020,iablokov2021} and used, e.g., in \cite{ayala2021,siri2024-1,siri2024-2,siri2024-3}, we then determine the free fermion propagator of this model in coordinate space. Then, carrying out an appropriate Fourier-Bessel transformation, we present it also in momentum space. These results appear for the first time in the literature.
\par
We then define the vector and axial vector currents $J^{\mu}_{V}$ and $J^{\mu}_{A}$ in terms of the free fermion propagator, expressed at finite temperature, and determine them as functions of temperature $T$, vector and axial chemical potential, $\mu$ and $\mu_{5}$, as well as the angular velocity $\Omega$. We show that in the first approximation, these currents are proportional to $\Omega$. The proportionality factors are the chiral vortical conductivities $\sigma_{V}$ and $\sigma_{A}$, corresponding to $J^{\mu}_{V}$ and $J^{\mu}_{A}$.  The final results for $\sigma_{V}$ and $\sigma_{A}$ coincide, as expected, with those in the literature \cite{kharzeev2015, landsteiner2011, landsteiner2013, liao}. Notably, the fact that the free fermion propagator developed in the first part of the paper yields the standard results for $J^{\mu}_{V}$ and $J^{\mu}_{A}$ confirms its validity. This is important, since these propagators are also used in different context in thermal field theory, e.g. in determining the thermodynamic potential of a charged, chiral, and rigidly rotating free Fermi gas which yields all thermodynamic quantities related to this gas. 
\par
In the second part of the paper, we introduce for the first time a method to separately compute the moment of inertia corresponding to the spin and orbital angular momentum in a rotating Fermi gas exhibiting a charge and chirality imbalance. We denote these moments of inertia by $I_{S}$ and $I_{L}$. To determine $I_{S}$, we utilize a certain relation between the axial vector current and the spin angular momentum density \cite{fukushima2018}. This enables us to use the result for $J_{A}^{\mu}$ from the first part of the paper to determine $I_{S}$. We then calculate the total moment of inertia, $I_{\text{tot}}$. We show that $I_{\text{tot}}=I_{S}$, implying that the orbital part of the moment of inertia vanishes ($I_{L}=I_{\text{tot}}-I_{S}=0$). The primary motivation for this computation was a recent work by Chernodub \textit{et al.} \cite{chernodub2023}, where the total moment of inertia of a gluonic gas at temperatures both below and above a specific supervortical temperature, $T_{s}$ is determined. Here, it is shown that at temperatures below (above) $T_{s}$, the moment of inertia of the spin-one gluon gas is negative (positive). A similar effect is also observed in \cite{siri2024-1}, where the moment of inertia of a spin-zero bosonic gas turns out to be negative under specific conditions. As it is supposed in \cite{chernodub2023}, a competition between the moment of inertia associated with the orbital and spin angular momenta leads under certain circumstances to a negative net moment of inertia and eventually to a negative Barnett effect.  Despite this indication, it has not yet been possible to calculate the moments of inertia $I_{S}$ and $I_{L}$ separately. We notice at this stage that a decomposition of the total angular momentum into orbital and spin contributions is theoretically affected by ambiguities arising from the possibility of redefining the spin part of the total angular momentum through pseudogauge transformation (see the recent review \cite{speranza2021} and the references therein). It can be shown that various pseudogauge transformations yield different spin currents, including canonical, Belifante-Rosenfeld (BR), Hilgevoord-Wouthuysen (HW), and de Groot-van Leeuwen-van Wert (GLW) currents. As it is argued in \cite{speranza2021}, HW and GLW pseudogauges naturally arise only in the \textit{massive case}, where it is possible to define the global spin in the particle rest frame. This is in contrast to the \textit{massless case}, for which a covariant definition of a global spin is not possible. As concerns the canonical and BR pseudogauges for massless fermions, they yield different results when we decompose the total angular momentum into its orbital and spin components. In the canonical pseudogauge the spin part of the total angular momentum is nonvanishing, whereas it vanishes in the BR pseudogauge. Hence, our results of $I_{S}$  and $I_{L}$ for massless fermions, which are determined using the canonical pseudogauge, are not only frame dependent, but also affected by the pseudogauge ambiguity. Despite these theoretical complexities, we find it valuable to present a method to separately compute the global spin angular momentum density and its corresponding moment of inertia for massless fermions.
\par
The organization of the paper is as follows: in Sec. \ref{sec2}, we introduce the Lagrangian density of rigidly rotating fermions in a charged and
chirally imbalanced medium. We derive the corresponding equation of motion and solve it. We use these solutions to compute the free propagator of these fermions by utilizing the Fock-Schwinger method. In Sec. \ref{sec3}, we define $J_{V}^{\mu}$ and $J_{A}^{\mu}$ in terms of this propagator and determine the chiral vortical conductivities $\sigma_{V}$ and $\sigma_{A}$. In Sec. \ref{sec4}, we then focus on the moment of inertia $I_{S}$ and $I_{L}$ and show that $I_{L}=0$. Section \ref{sec5} is devoted to our concluding remarks. Appendices \ref{appA} and \ref{appB} present useful mathematical relations. Appendix \ref{appC} includes the derivation of three relations. In Appendix \ref{appD}, apart from a certain derivation, we describe the relation between the canonical form of the spin tensor with the spin angular momentum density defined in Sec. \ref{sec4}.
\section{Rotating fermions in a charged and chirally imbalanced medium}\label{sec2}
\setcounter{equation}{0}
\subsection{The model}\label{sec2A}
Let us start with the Lagrangian density
\begin{eqnarray}\label{A1}
	\mathscr{L} = \bar{\psi} \big[i\gamma^{\alpha}\nabla_{\alpha} -m +\mu\gamma^{0}+\mu_{5}\gamma^{0}\gamma^{5} \big]\psi,
\end{eqnarray}
for a free fermion field $\psi$, with mass $m$ in a charged and chirally imbalanced medium characterized by the vector and axial chemical potentials $\mu$ and $\mu_{5}$. Assuming a rigid rotation, the covariant derivative in \eqref{A1} is defined by
\begin{eqnarray}\label{A2}
\nabla_{\mu} \psi = (\partial_{\mu} + \Gamma_{\mu} ) \psi,
\end{eqnarray}
with the spin connection $\Gamma_{\mu}$ given by
$\Gamma_{\mu} = -\frac{i}{4} \omega_{\mu i j} \sigma^{ij}
$. Here,
$\omega_{\mu i j}\equiv g_{\alpha \beta} e^{\alpha}_{\,i} (\partial_{\mu} e^{\beta}_{\, j} + \Gamma^{\beta}_{\mu\nu} e^{\nu}_{\, j})$ and $\sigma^{ij} = \frac{i}{2} [\gamma^{i} , \gamma^{j}]$. In $\omega_{\mu i j}$, the tetrads $e_{i}^{\alpha}$, defined by $\eta_{ij}=e_{i}^{\alpha}e_{j}^{\beta}g_{\alpha\beta}$ and the Christoffel symbol, defined by $\Gamma_{\mu\nu}^{\lambda}=\frac{1}{2}g^{\lambda \sigma} (\partial_{\mu}g_{\sigma \nu} + \partial_{\nu}g_{\sigma \mu} -\partial_{\sigma} g_{\mu \nu} )$, are expressed in terms of the metric $g_{\mu\nu}$, that for a rigid rotation around the $z$ axis with the angular velocity $\bs{\Omega}=\Omega\bs{e}_{z}$ reads
\begin{equation}\label{A3}
	g_{\mu\nu} =
	\begin{pmatrix}
		1-(x^{2}+y^{2})\Omega^{2} & \Omega y & -\Omega x & 0\\
		\Omega y & -1 & 0 & 0 \\
		-\Omega x & 0 & -1 & 0 \\
		0 & 0 & 0 & -1
	\end{pmatrix}  .
\end{equation}
Here, $x$ and $y$ are Cartesian coordinates. Let us notice that the Greek and Latin indices $\alpha,\beta\in \{t,x,y,z\}$ and $i,j\in \{0,1,2,3\}$ are the corresponding spacetime indices for the corotating observer and the laboratory (inertial) frames, respectively. In addition, Dirac $\gamma$ matrices $\gamma^{\alpha}$ in the rotating frame are related to $\gamma^{i}$ in the inertial frame by $\gamma^{\mu}=e^{\mu}_{i}\gamma^{i}$, where $\gamma^{i}$s satisfy $\{\gamma^{i},\gamma^{j}\}=2\eta^{ij}$. Using $g_{\mu\nu}$ from \eqref{A3}, the nonvanishing components of $\Gamma_{\mu\nu}^{\lambda}$ in the Cartesian coordinates are
\begin{eqnarray}\label{A4}
\Gamma^{x}_{tt} = -x\Omega^{2}  ,&\qquad& \Gamma^{y}_{tt} = -y\Omega^{2}  ,\nonumber\\
\Gamma^{y}_{tx} = \Gamma^{y}_{xt} = \Omega   ,&\qquad& \Gamma^{x}_{ty} = \Gamma^{x}_{yt} = -\Omega.
\end{eqnarray}
Choosing
\begin{eqnarray}\label{A5}
e^{t}_{0} = e^{x}_{1} = e^{y}_{2} = e^{z}_{3} = 1, \quad e^{t}_{1} = y\Omega, \quad  e^{t}_{2} = - x \Omega,\nonumber\\
\end{eqnarray}
we arrive at
\begin{eqnarray}\label{A6}
	\omega_{t12} = - \omega_{t21} = \Omega,
\end{eqnarray}
leading to $	\Gamma_{t} = - \frac{i}{2} \Omega \sigma^{12}  $. All the other components of the spin connection vanish. The Dirac $\gamma$ matrices are then given by
\begin{eqnarray}\label{A7}
		\gamma^{t} = \gamma^{0} ,&\qquad&		\gamma^{z} = \gamma^{3} ,\nonumber\\
		\gamma^{x} = \gamma^{1} + y\Omega \gamma^{0} ,&\qquad&
		\gamma^{y} = \gamma^{2} - x\Omega \gamma^{0}.
\end{eqnarray}
Moreover, using
\begin{eqnarray}\label{A8}
	\gamma_{5} = -i \frac{\sqrt{-g}}{4!} \epsilon_{\alpha\beta\mu\nu} \gamma^{\alpha}\gamma^{\beta}\gamma^{\mu}\gamma^{\nu},
\end{eqnarray}
with $g\equiv \text{det}(g_{\mu\nu})=-1$,
it turns out that $\gamma^{5}$ is invariant under rigid rotation. In this paper, we use $\gamma^{5}=i\gamma^{0}\gamma^{1}\gamma^{2}\gamma^{3}$. Combining these results, we arrive at the Dirac equation of free fermions in a rotating medium with finite $\mu$ and $\mu_{5}$,
\begin{eqnarray}\label{A9}
	\big[  \gamma^{0} \left( i\partial_{t} +\Omega J_{z} + \mu \right) + i\bs{\gamma}\cdot \bs{\nabla} +\mu_{5}\gamma^{0}\gamma^{5}-m \big] \psi = 0.\nonumber\\
\end{eqnarray}
Here, the total angular momentum in the $z$ direction reads $J_{z}\equiv L_{z}+S_{z}$, with the orbital angular momentum $L_{z}\equiv -i\left(x\partial_{y}-y\partial_{x}\right)$ and the spin $S_{z}\equiv \sigma^{12}/2$ with $\sigma^{12}=\mathbb{I}_{2\times 2}\otimes \sigma^{3}$. In this paper, we use the Dirac representation for the Dirac $\gamma$ matrices,
\begin{eqnarray}\label{A10}
\gamma^{0}=	\begin{pmatrix}
		0 & 1\\
1 & 0
	\end{pmatrix},\qquad
\bs{\gamma}=	\begin{pmatrix}
		0 & \bs{\sigma}\\
-\bs{\sigma} & 0
	\end{pmatrix},
\end{eqnarray}
with $\mathbb{I}_{2\times 2}=\text{diag}(1,1)$ and $\bs{\sigma}=\left(\sigma^{1},\sigma^{2},\sigma^{3}\right)$ the Pauli matrices
satisfying $[\sigma^{i},\sigma^{j}]=2i\epsilon^{ijk}\sigma^{k}$. In what follows, we present the solution of the Dirac equation \eqref{A9}.
To this purpose, we use cylindrical coordinate system $x^{\mu}=\left(t,x,y,z\right)=(t,\rho \cos\varphi,\rho\sin\varphi,z)$, with $\rho$ the
radial coordinate, $\varphi$ the azimuthal angle, and $z$ the height of the cylinder.
\subsection{Solution of Dirac equation \eqref{A9}}\label{sec2B}
To solve \eqref{A9}, we first consider the eigenvalue equation for $J_{z}$,
\begin{eqnarray}\label{A11}
J_{z}\psi=j\psi,
\end{eqnarray}
whose eigenvalue turns out to be $j=\ell+1/2$. Having in mind that in the cylindrical coordinate system the orbital angular momentum is given by $L_{z}=-i\partial_{\varphi}$, it is advantageous to use $e^{i\ell\varphi}$ and $e^{i(\ell+1)\varphi}$ as the eigenvalues of $J_{z}$. Inspired by the method used in \cite{ayala2021}, we use the ansatz
\begin{eqnarray}\label{A12}
\psi=
\big[  \gamma^{0} \left( i\partial_{t} +\Omega J_{z}+ \mu \right) + i\bs{\gamma}\cdot \bs{\nabla}+\mu_{5}\gamma^{0}\gamma^{5} +m\big] \Phi, \nonumber\\
\end{eqnarray}
that, upon inserting into \eqref{A9}, leads to a second-order differential equation for $\Phi$
\begin{eqnarray}\label{A13}
	\Big[ \Big( i\partial_{t} +\Omega J_{z} +\mu \Big)^{2}  + \bs{\nabla}^{2} -\mu_{5}^{2} -4i\mu_{5} \bs{S}\cdot\bs{\nabla}-m^{2} \Big] \Phi = 0.\nonumber\\
\end{eqnarray}
Here, the spin operator $\bs{S}$ is defined by
\begin{eqnarray}\label{A14}
	\bs{S} \equiv \frac12 \gamma^{5}\gamma^{0}\bs{\gamma} =\mbox{diag}\left(\bs{\sigma},\bs{\sigma}\right).
\end{eqnarray}
In what follows, we first determine the solutions of the Klein-Gordon-like differential equation \eqref{A13}, which yields the solution of the Dirac equation \eqref{A9} $\psi$ upon inserting into \eqref{A12}.
\par
To solve \eqref{A13}, we use the helicity operator $\mathfrak{h}=\bs{S}\cdot \bs{\hat{k}}$, with $\bs{\hat{k}}\equiv \bs{k}/|\bs{k}|$ and $\bs{k}=-i\bs{\nabla}$ being the spatial components of a four-momentum $k$. Using the fact that $\mathfrak{h}$ commutes with $J_{z}$ and the differential operator $\Big[ \Big( i\partial_{t} +\Omega J_{z} +\mu \Big)^{2}  + \bs{\nabla}^{2} -\mu_{5}^{2} -4i\mu_{5} \bs{S}\cdot\bs{\nabla}-m^{2} \Big]$, arising in \eqref{A13}, the components of $\Phi$ can be chosen to be the eigenstates of the helicity operator $\mathfrak{h}$ with eigenvalues $\lambda/2$ with $\lambda=\pm 1$. This justifies the following ansatz for $\Phi$:
\begin{eqnarray}\label{A15}
\Phi_{\eta}= \left\{
\begin{array}{rcl}
\Phi_{1,\eta}&=&
		\begin{pmatrix}
			\phi_{\eta}(x,k) \\ \bs{0}
		\end{pmatrix},\\
\\
\Phi_{2,\eta}&=&
		\begin{pmatrix}
			\bs{0}\\ \phi_{\eta}(x,k)
		\end{pmatrix},
\end{array}
\right.
\end{eqnarray}
with $\eta=\{\ell,\lambda\}$, $x=(t,\rho,\varphi,z)$, $\bs{0}$ is a $2\times 1$ null vector, and
\begin{eqnarray}\label{A16}
\hspace{-0.5cm}\phi_{\eta}(x,k)=\begin{pmatrix}
A_{\lambda}e^{i\ell\varphi}J_{\ell}(k_{\perp}\rho)\\
B_{\lambda}e^{i(\ell+1)\varphi}J_{\ell+1}\left(k_{\perp}\rho\right)
\end{pmatrix}e^{-iEt+ik_{z}z}.
\end{eqnarray}
Here, $J_{\ell}(z)$ is the Bessel function of order $\ell$ and $k_{\perp}=\left(k_x^{2}+k_y^{2}\right)^{1/2}$. To determine the normalization factors $A_{\lambda}$ and $B_{\lambda}$, we demand that $\phi_{\eta}$ is the eigenstate of the helicity operator corresponding to the eigenvalue $\lambda$ and arrive at
\begin{eqnarray}\label{A17}
\mathcal{M} \begin{pmatrix}
		A_{\lambda} \\
		B_{\lambda}
	\end{pmatrix} = 0,
\end{eqnarray}
with
\begin{eqnarray}\label{A18}
\mathcal{M}\equiv	\begin{pmatrix}
		k_{z} - \lambda |\bs{k}| & -ik_{\perp} \\
		ik_{\perp} & -k_{z} - \lambda |\bs{k}|
	\end{pmatrix}.
\end{eqnarray}
This leads immediately to
\begin{eqnarray}\label{A19}
\left(k_{z} - \lambda |\bs{k}|\right)A_{\lambda}-ik_{\perp}B_{\lambda}=0.
\end{eqnarray}
Moreover, to obtain nontrivial solutions for $A_{\lambda}$ and $B_{\lambda}$, we set $\text{det}(\mathcal{M})=0$ and arrive at
$|\boldsymbol{k}|=\left(k_{z}^{2}+k_{\perp}\right)^{1/2}$. An appropriate choice for $A$ and $B$ is
	\begin{eqnarray}\label{A20}
		A_{\lambda} &=& \frac1{\sqrt{2}} \sqrt{1+\lambda \frac{k_{z}}{|\bs{k}|}} ,\nonumber\\
		B_{\lambda}&=&\frac{i\lambda}{\sqrt{2}} \sqrt{1-\lambda \frac{k_{z}}{|\bs{k}|}}.
	\end{eqnarray}
Another choice is presented recently in \cite{hua2024}.
Plugging at this stage \eqref{A16} into \eqref{A13}, we obtain the energy dispersion relation
	\begin{eqnarray}\label{A21}
		E\equiv \tilde{E}-\mu_{\text{eff}},
	\end{eqnarray}
where $\tilde{E}^{2} \equiv  (|\bs{k}|-\lambda \mu_{5})^{2}+m^{2}$, $\mu_{\text{eff}}\equiv j\Omega+\mu$ and $j=\ell+1/2$. Setting $\Omega=0$ and $\mu=0$, the above dispersion relation is the same as appears in \cite{fukushima2009}. Let us notice that for $\mu=0$, the energy in the corotating frame $E$ is given by $E=\tilde{E}-\bs{J}\cdot\bs{\Omega}$, where $\tilde{E}$ is the energy in the laboratory frame and $\bs{J}$ is the total angular momentum \cite{landau-book}.
\par
It turns out that the solution $\Phi$ from \eqref{A15} satisfies the following orthonormality:
\begin{eqnarray}\label{A22}
\sum_{\ell=-\infty}^{\infty}\Phi_{i,\{\ell,\lambda\}}^{\dagger}(x,k)\Phi_{j,\{\ell,\lambda^{\prime}\}}(x,k)=\delta_{ij}\delta_{\lambda\lambda^{\prime}},
\end{eqnarray}
and closure relations
\begin{eqnarray}\label{A23}
\sum_{\kappa}\Phi_{\kappa}(x,k)\Phi^{\dagger}_{\kappa}(x',k)= I_{4\times4} \delta^{4}(x-x^{\prime}).
\end{eqnarray}
Here, we introduced the notation
\begin{eqnarray}\label{A24}
\sum_{\kappa}\equiv \sum_{\ell=-\infty}^{\infty}\sum_{\lambda=\pm 1}\sum_{i=1,2}\int\frac{dEdk_{z}dk_{\perp}k_{\perp}}{(2\pi)^{3}},
\end{eqnarray}
with $\{E,k_{z}\}\in(-\infty,\infty)$  and $k_{\perp}\in [0,\infty)$.
\par
Substituting, at this stage, the solutions \eqref{A15} and \eqref{A16} for $\Phi$ into \eqref{A13}, the solution $\psi$ for the Dirac equation reads
\begin{eqnarray}\label{A26}	\psi(x,k)&=& \begin{pmatrix}
			\Big( \mathcal{E}^{-}A_{\lambda}+ik_{\perp}B_{\lambda} \Big) J_{\ell}(u) \\
			\Big( \mathcal{E}^{+}B_{\lambda} -ik_{\perp}A_{\lambda}\Big) e^{i\varphi}J_{\ell+1}(u) \\
			-\Big(\mathcal{F}^{-}A_{\lambda} +ik_{\perp}B_{\lambda} \Big) J_{\ell}(u) \\
			-\Big(\mathcal{F}^{+}B_{\lambda} -ik_{\perp}A_{\lambda} \Big)e^{i\varphi}J_{\ell+1}(u)
		\end{pmatrix} \nonumber\\
&&\times e^{-iEt+i\ell\varphi+ik_{z}z},
\end{eqnarray}
with
$\mathcal{E}^{\pm}=\tilde{E}+ m+\mu_{5}\pm k_{z}$, $\mathcal{F}^{\pm}=\tilde{E}-m+\mu_{5}\pm k_{z}$, $u\equiv k_{\perp}\rho$,
and $\tilde{E}=E+\mu_\text{eff}$.
\subsection{Fock-Schwinger proper-time method and free fermion propagator}\label{sec2C}
As it is argued in the previous section, the fermion wave function $\psi$ from \eqref{A26} is directly related to the bosonic field $\Phi$ appearing in \eqref{A12}, which satisfies \eqref{A13}. It is, therefore, possible to use the standard relation
\begin{eqnarray}\label{A27}
S(x,x^{\prime}) &=& [\gamma^{0} \left( i\partial_{t} +\Omega J_{z} + \mu \right) + i\bs{\gamma}\cdot \bs{\nabla} \nonumber\\
&&+\mu_{5}\gamma^{0}\gamma^{5} +m] G(x,x^{\prime}),
\end{eqnarray}
between the free fermion propagator $S(x,x')$ and the free boson propagator $G(x,x')$ to determine $S(x,x')$. In what follows, we determine $G(x,x')$ by utilizing the Fock-Schwinger method \cite{iablokov2019,iablokov2020,iablokov2021,ayala2021,siri2024-1,siri2024-2}. Plugging the resulting expression for $G(x,x')$ into \eqref{A27} leads eventually to the fermion propagator $S(x,x')$.
\par
To determine $G(x,x')$, let us begin with Green's equation
\begin{eqnarray}\label{A28}
	\mathcal{D}(\partial_{x} ,x) \mathcal{G}(x, x^{\prime}) = \delta^{4}(x-x^{\prime}),
\end{eqnarray}
with $\mathcal{D}(\partial_{x} ,x)$ a generic differential operator. The Green function  $\mathcal{G}(x,x^{\prime})$ is represented as
\begin{eqnarray}\label{A29}
\mathcal{G}(x,x^{\prime})=-i\int_{-\infty}^{0}U(x,x^{\prime};\tau) d\tau,
\end{eqnarray}
where $U(x,x^{\prime};\tau)$ is the proper-time evolution operator and $\tau$ the proper time. The former satisfies a Schr\"odinger-like equation
\begin{eqnarray}\label{A30}
	i\partial_{\tau} U(x,x^{\prime}; \tau) = \mathcal{D}(\partial_{x} , x) U(x,x^{\prime}; \tau).
\end{eqnarray}
With appropriate boundary conditions
\begin{eqnarray}\label{A31}
&&\lim\limits_{\tau\to 0}U(x,x^{\prime};\tau)=\delta^{4}(x-x'),\nonumber\\
&&\lim\limits_{\tau\to -\infty}U(x,x';\tau)=0,
\end{eqnarray}
the solution of \eqref{A30} reads
\begin{eqnarray}\label{A32}
	U(x,x^{\prime}; \tau) = e^{-i\tau \mathcal{D}(\partial_{x} , x)+\epsilon\tau} \delta^{4}(x-x^{\prime}) .
\end{eqnarray}
Here, $\epsilon>0$ is necessary to guarantee that $U(x,x';\tau)$ vanishes at $\tau\to -\infty$ [see the second condition in \eqref{A31}]. We will omit the factor $\epsilon$ in the rest of this paper.\footnote{If we keep $\epsilon$ and follow the steps leading from \eqref{A37} to free fermion propagator \eqref{A42}, an additional term $i\epsilon$ will appear, as expected, in the denominator of fermion propagators. Having this in mind, we will omit $\epsilon$ in these steps, especially because in the following sections, we will work with fermion propagators at finite temperature.}
\par
At this stage, we consider the eigenvalue equation of the differential operator $\mathcal{D}(\partial_{x} ,x)$ in \eqref{A28},
\begin{eqnarray}\label{A33}
\mathcal{D}(\partial_{x},x)\varphi_{\sigma}(x)=\sigma\varphi_{\sigma}(x),
\end{eqnarray}
where $\varphi_{\sigma}$ and $\sigma$ are the eigenfunction and eigenvalue of $\mathcal{D}(\partial_{x},x)$, respectively. Using the fact that the eigenfunctions $\varphi_{\sigma}(x)$ satisfy the closure relation
\begin{eqnarray}\label{A34}
	\sum_{\sigma} \varphi_{\sigma}(x) \varphi^{\dagger}_{\sigma}(x^{\prime}) = \delta^{4}(x-x^{\prime}),
\end{eqnarray}
the proper-time evolution operator $U(x,x^{\prime};\tau)$ is thus given by
\begin{eqnarray}\label{A35}
	U(x,x^{\prime}; \tau) = \sum_{\sigma} e^{-i\tau \sigma} \varphi_{\sigma}(x) \varphi^{\dagger}_{\sigma}(x^{\prime}).
\end{eqnarray}
Plugging \eqref{A35} into \eqref{A29}, the free bosonic Green function $\mathcal{G}(x,x')$ reads
\begin{equation}\label{A36}
	\mathcal{G}(x,x^{\prime}) = -i \int_{-\infty}^{0} d\tau \sum_{\sigma} e^{-i\tau \sigma} \varphi_{\sigma}(x) \varphi^{\dagger}_{\sigma}(x^{\prime}).
\end{equation}
Let us now consider the differential equation \eqref{A13} that yields the Green's function equation
\begin{eqnarray}\label{A37}
\mathscr{D}G(x,x')=\delta^{4}(x-x'),
\end{eqnarray}
with the differential operator
\begin{eqnarray}\label{A38}
\mathscr{D} &\equiv&  \Big( i\partial_{t} +\Omega J_{z} +\mu \Big)^{2}  + \bs{\nabla}^{2} -\mu_{5}^{2} -4i\mu_{5} \bs{S}\cdot\bs{\nabla}-m^{2}.\nonumber\\
\end{eqnarray}
Following the above steps,  the Green function $G(x,x')$ from \eqref{A37} is first given by
\begin{eqnarray}\label{A39}
\hspace{-1cm}G(x,x^{\prime}) = -i \int_{-\infty}^{0} d\tau e^{-i\tau \mathscr{D}} I_{4\times4}\delta^{4}(x-x^{\prime}).
\end{eqnarray}
Then, using the closure relation \eqref{A23}, it is possible to replace $ I_{4\times4}\delta^{4}(x-x^{\prime})$ on the rhs of \eqref{A39} with $\sum_{\kappa}\Phi_{\kappa}(x,k)\Phi_{\kappa}^{\dagger}(x',k)$. Plugging $\Phi_{\kappa}$ from \eqref{A15} and \eqref{A16} into this expression, we arrive first at
\begin{eqnarray}\label{A25}
\lefteqn{\sum_{\kappa}\Phi_{\kappa}(x,k)\Phi^{\dagger}_{\kappa}(x',k)=\frac{1}{2}\sum_{\ell=-\infty}^{\infty}\sum_{\lambda=\pm 1}\int\frac{dEdk_{z}dk_{\perp}k_{\perp}}{(2\pi)^{3}}
}\nonumber\\
&&\times e^{-iE(t-t^{\prime}) + ik_{z}(z-z^{\prime}) + i\ell(\varphi-\varphi^{\prime})}
\nonumber \\
	&&\times \Bigg[ \left( 1+\lambda \frac{k_{z}}{|\bs{k}|} \gamma^{5}\gamma^{0}\gamma^{3} \right) \left( \mathcal{O}_{+} \mathcal{J}_{\ell,\ell} + \mathcal{O}_{-} \mathcal{J}_{\ell+1,\ell+1}e^{+i(\varphi-\varphi^{\prime})} \right)\nonumber\\
&& +\lambda\frac{k_{\perp}}{|\bs{k}|}\gamma^{5}\gamma^{0}\gamma^{2} 	 \left( \mathcal{O}_{+}\mathcal{J}_{\ell+1,\ell}e^{i\varphi}  + \mathcal{O}_{-}\mathcal{J}_{\ell,\ell+1}e^{-i\varphi^{\prime}} \right)  \Bigg],
\end{eqnarray}
where $\mathcal{O}_{\pm}\equiv\left(1\pm i\gamma^{1}\gamma^{2}\right)/2$, and $\mathcal{J}_{L,L'}\equiv J_{L}(k_{\perp}\rho)J_{L'}(k_{\perp}\rho')$.
Eventually using the eigenvalue equation
\begin{eqnarray}\label{A40}
	\mathscr{D}\Phi_{\kappa}(x,k) = \Big(\tilde{E}^{2} -\omega_{\lambda}^{2} \Big)\Phi_{\kappa}(x,k),
\end{eqnarray}
with $\omega_{\lambda}^{2}\equiv (|\bs{k}|-\lambda \mu_{5})^{2}+m^{2}$, we arrive after integrating over the proper time $\tau$ at
\begin{widetext}
\begin{eqnarray}\label{A41}
G(x,x^{\prime})&=& \frac{1}{2} \sum_{\ell=-\infty}^{\infty}\sum_{\lambda=\pm}\int\frac{dEdk_{z}dk_{\perp}k_{\perp}}{(2\pi)^{3}}\frac{ e^{-iE(t-t^{\prime}) + ik_{z}(z-z^{\prime}) + i\ell(\varphi-\varphi^{\prime})} }{\tilde{E}^{2} - \omega_{\lambda}^{2}} \Bigg[ \left( 1+\lambda \frac{k_{z}}{|\bs{k}|} \gamma^{5}\gamma^{0}\gamma^{3} \right)
 \nonumber \\
	&&\times\left( \mathcal{O}_{+} \mathcal{J}_{\ell,\ell} + \mathcal{O}_{-} \mathcal{J}_{\ell+1,\ell+1}e^{+i(\varphi-\varphi^{\prime})} \right)+\lambda\frac{k_{\perp}}{|\bs{k}|}\gamma^{5}\gamma^{0}\gamma^{2} 	 \left( \mathcal{O}_{+}\mathcal{J}_{\ell+1,\ell}e^{i\varphi}  + \mathcal{O}_{-}\mathcal{J}_{\ell,\ell+1}e^{-i\varphi^{\prime}}\right)  \Bigg],
\end{eqnarray}
where $\tilde{E}=E+\mu_{\text{eff}}$.
Plugging, then, the above result into \eqref{A27}, the free fermion propagator in the coordinate space is given by
\begin{eqnarray}\label{A42}
	S(x,x^{\prime})&=&\frac{1}{2}\sum_{\ell=-\infty}^{\infty}\sum_{\lambda=\pm}\int\frac{dEdk_{z}dk_{\perp}k_{\perp}}{(2\pi)^{3}} \frac{ e^{-iE(t-t^{\prime}) + ik_{z}(z-z^{\prime}) + i\ell(\varphi-\varphi^{\prime})} }{\tilde{E}^{2} -\omega_{\lambda}^{2}} \nonumber \\
	&&\times \Bigg\{ \left(\tilde{E}\gamma^{0}+m+\mu_{5}\gamma^{0}\gamma^{5} - k_{z}\gamma^{3} \right) \Big[\left( 1+\lambda \frac{k_{z}}{|\bs{k}|} \gamma^{5}\gamma^{0}\gamma^{3} \right) \left( \mathcal{O}_{+} \mathcal{J}_{\ell,\ell} + \mathcal{O}_{-} \mathcal{J}_{\ell+1,\ell+1}e^{+i(\varphi-\varphi^{\prime})} \right) \nonumber \\
&&\qquad +\lambda\frac{k_{\perp}}{|\bs{k}|}\gamma^{5}\gamma^{0}\gamma^{2} 	 \left( \mathcal{O}_{+}\mathcal{J}_{\ell+1,\ell}e^{i\varphi}  + \mathcal{O}_{-}\mathcal{J}_{\ell,\ell+1}e^{-i\varphi^{\prime}} \right)  \Big] + \lambda k_{\perp} \frac{k_{\perp}}{|\bs{k}|}\gamma^{5}\gamma^{0}\left( \mathcal{O}_{+} \mathcal{J}_{\ell,\ell} + \mathcal{O}_{-} \mathcal{J}_{\ell+1,\ell+1}e^{+i(\varphi-\varphi^{\prime})} \right) \nonumber \\
	&&\qquad -k_{\perp}\gamma^{2}\left( 1+\lambda \frac{k_{z}}{|\bs{k}|} \gamma^{5}\gamma^{0}\gamma^{3} \right)\left( \mathcal{O}_{+}\mathcal{J}_{\ell+1,\ell}e^{i\varphi}  + \mathcal{O}_{-}\mathcal{J}_{\ell,\ell+1}e^{-i\varphi^{\prime}} \right)  \Bigg\}.
\end{eqnarray}
\end{widetext}
Let us notice that the projector operators $\mathcal{O}_{\pm}$ appearing in the above expressions
 satisfy $\mathcal{O}_{\pm}\mathcal{O}_{\pm}=\mathcal{O}_{\pm}$, $\mathcal{O}_{\pm}\mathcal{O}_{\mp}=0$, and $\mathcal{O}_{+}+\mathcal{O}_{-}=1$. For $\gamma^{\mu_{\|}}\equiv \left(\gamma^{0},0,0,\gamma^{3}\right)$ and $\gamma^{\mu_{\perp}}\equiv \left(0,\gamma^{1},\gamma^{2},0\right)$, we have
$\mathcal{O}_{\pm}\gamma^{\mu_{\|}}=\gamma^{\mu_{\|}}\mathcal{O}_{\pm}, \mathcal{O}_{\pm}\gamma^{\mu_{\perp}}=\gamma^{\mu_{\perp}}\mathcal{O}_{\mp}$. Moreover, $\mathcal{O}_{\pm}\gamma^{5}=\gamma^{5}\mathcal{O}_{\pm}$, and $\gamma^{1}\mathcal{O}_{\pm}=\mp i\gamma^{2}\mathcal{O}_{\pm}$.
\par
To determine the above fermion propagator in momentum space, we perform an appropriate inverse Fourier-Bessel transformation of $S(x,x')$ in the cylindrical coordinate system,
\begin{eqnarray}\label{A43}
\hspace{-1cm}S_{\ell\ell'}(p,p')=\int d^{4}xd^{4}x' S(x,x')\phi_{\ell}^{\dagger}(x,p)\phi_{\ell'}(x',p'),
\end{eqnarray}
with $d^{4}x=dtd\varphi dz d\rho$ and
\begin{eqnarray}\label{A44}
\phi_{\ell}(x,p)\equiv e^{-ip_{0}t+ip_{z}z+i\ell\varphi}J_{\ell}\left(p_{\perp}\rho\right).
\end{eqnarray}
Using the orthonormality relations presented in Appendix \ref{appA}, we arrive at
\begin{eqnarray}\label{A45}
S_{\ell\ell'}(p,p')= (2\pi)^{3}\widehat{\delta}_{\ell,\ell'}^{3}\left(p_0,p_z,p_{\perp};p_{0}^{\prime},p_{z}^{\prime},p_{\perp}^{\prime}\right)S_{\ell}(p),\nonumber\\
\end{eqnarray}
with
\begin{eqnarray}\label{A46}
\widehat{\delta}_{\ell,\ell'}^{3}\left(p_0,p_z,p_{\perp};p_{0}^{\prime},p_{z}^{\prime},p_{\perp}^{\prime}\right)&\equiv&\frac{1}{p_{\perp}} \delta(p_{0}-p^{\prime}_{0})\delta(p_{z}-p^{\prime}_{z})\nonumber\\
&&\times  \delta(p_{\perp}-p^{\prime}_{\perp})\delta_{\ell,\ell^{\prime}},\nonumber\\
\end{eqnarray}
and
\begin{eqnarray}\label{A47}
S_{\ell}(p)&=&\sum_{\lambda=\pm 1}\Big[  \frac{E_{+}\gamma^{0}+m+\mu_{5}\gamma^{0}\gamma^{5}-p_{z}\gamma^{3}-p_{\perp}\gamma^{2}}{E_{+}^{2} -\omega_{\lambda}^{2}} \Gamma_{\lambda} \mathcal{O}_{+}\nonumber\\
&& + \frac{E_{-}\gamma^{0}+m+\mu_{5}\gamma^{0}\gamma^{5}-p_{z}\gamma^{3}-p_{\perp}\gamma^{2}}{E_{-}^{2} -\omega_{\lambda}^{2}} \Gamma_{\lambda} \mathcal{O}_{-} \Big]. \nonumber\\
\end{eqnarray}
Here, $E_{\pm}\equiv p_{0}+\left(\ell\pm 1/2\right)+\mu$ and
$$\Gamma_{\lambda} \equiv \frac12\left(1+ \lambda \frac{p_{z}}{|\bs{p}|} \gamma^{5}\gamma^{0}\gamma^{3}+\lambda\frac{p_{\perp}}{|\bs{p}|}\gamma^{5}\gamma^{0}\gamma^{2}\right).$$
Let us notice that $\Gamma_{\lambda}$ is the cylindrical version of the projection operator
$	\Gamma_{\lambda}(\bs{p}) = \frac{1}{2} \left(1+\lambda\frac{\bs{p}\cdot\bs{\gamma}}{|\bs{p}|}\gamma^{0}\gamma^{5} \right)$,
defined in \cite{fukushima2009}. Replacing $p_0$ in $E_{\pm}$ with fermionic Matsubara frequencies $i\omega_{n}\equiv i(2n+1)\pi T$ and $n\in \mathbb{Z}$, we obtain the fermion propagator at finite temperature $T$. To arrive at \eqref{A47}, we used the same method as described explicitly in \cite{siri2024-1}.
\section{Chiral Vortical Conductivities}\label{sec3}
\setcounter{equation}{0}
In this section, we study the chiral vortical effect by determining the third component of the vector and axial vector currents, $J_{V}^{\mu} $ and $J_{A}^{\mu}$ in the chiral limit $m=0$. They are determined by
\begin{eqnarray}\label{B1}
J^{\|}_{V} = -i \lim_{x^{\prime} \rightarrow x} \text{tr}(\gamma^{3} S(x,x^{\prime})),
\end{eqnarray}
and
\begin{eqnarray}\label{B2}
J^{\|}_{A} = -i \lim_{x^{\prime} \rightarrow x} \text{tr}(\gamma^{3}\gamma^{5} S(x,x^{\prime})),
\end{eqnarray}
where $S(x,x^{\prime})$ is given in \eqref{A42}. Here, $J^{\|}_{V}$ and $J^{\|}_{A}$ are aligned in the direction of angular velocity $\bs{\Omega}=\Omega\bs{e}_{z}$. To build the summation over $\ell$, we use the formulae presented in Appendix \ref{appB}.
\subsection{Vector current}\label{sec3A}
Plugging \eqref{A42} into \eqref{B1}, and evaluating the traces of Dirac $\gamma$ matrices, we arrive first at
\begin{eqnarray}\label{B3}
J^{\parallel}_{V}(x) &=& -i \sum_{\ell=-\infty}^{\infty}\sum_{\lambda=\pm 1}\int \frac{dEdk_{z}dk_{\perp}k_{\perp}}{(2\pi)^{3}} \frac{\lambda\omega_{\lambda}}{\left(\tilde{E}^{2}-\omega_{\lambda}^{2}\right)} \nonumber\\
&&\times [J^{2}_{\ell}(k_{\perp}\rho) - J^{2}_{\ell+1}(k_{\perp}\rho)],
\end{eqnarray}
with $\tilde{E}=E+\mu_{\text{eff}}$ and $\omega_{\lambda}=|\bs{k}|-\lambda\mu_{5}$ in the chiral limit. Then, using
\begin{eqnarray}\label{B4}
\int dE f(E)=2\pi iT\sum_{n=-\infty}^{\infty}f(i\omega_{n}),
\end{eqnarray}
with $\omega_{n}=(2n+1)\pi T$ and performing a change of variable $k_{\perp}\to |\bs{k}|$ using $k_{\perp}^{2}=|\bs{k}|^{2}-k_{z}^{2}$, the corresponding expression at finite temperature reads
\begin{eqnarray}\label{B5}
J^{\parallel}_{V} &=& -\frac{T}{(2\pi)^{2}}\sum_{\lambda=\pm 1} \sum_{\ell=-\infty}^{\infty}\sum_{n=-\infty}^{\infty} \nonumber\\
&&\times \int_{k}
 \frac{\lambda\omega_{\lambda}}{[(\omega_{n} -i\mu_{\text{eff}})^{2}+\omega_{\lambda}^{2}]} \big[J^{2}_{\ell}(k_{\perp}\rho) - J^{2}_{\ell+1}(k_{\perp}\rho)\big], \nonumber\\
\end{eqnarray}
with
\begin{eqnarray}\label{B6}
\int_{k}\equiv \int_{0}^{\infty} d|\bs{k}||\bs{k}| \int_{-|\bs{k}|}^{|\bs{k}|} dk_{z}.
\end{eqnarray}
Performing at this stage  the summation over Matsubara frequencies $\omega_{n}$ by utilizing
\begin{eqnarray}\label{B7}
	\sum_{n=-\infty}^{\infty} \frac{1}{\left( \omega_{n} -i\mu_{\text{eff}} \right)^{2} + \omega_{\lambda}^{2}} = \sum_{\epsilon=\pm}\frac{\tanh\left(\beta(\omega_{\lambda}+\epsilon \mu_{\text{eff}})/2\right)}{4\omega_{\lambda} T} ,\nonumber\\
\end{eqnarray}
with $\tanh(x)=1-2n_{\text{f}}(2x)$ and $n_{\text{f}}(x)\equiv (e^{x}+1)^{-1}$ the Fermi-Dirac distribution function, $J_{V}^{\|}$ is given by
\begin{equation}\label{B8}
J^{\|}_{V} = J^{\|}_{V,\text{vac}}+J^{\|}_{V,\text{mat}},
\end{equation}
while the vacuum ($T$-independent) part is given by
\begin{eqnarray}\label{B9}
J^{\|}_{V,\text{vac}} = \frac{-1}{2(2\pi)^{2}} \sum_{\lambda=\pm 1} \lambda \sum_{\ell=-\infty}^{\infty}\int_{k}\big[ J^{2}_{\ell}(k_{\perp}\rho) - J^{2}_{\ell+1}(k_{\perp}\rho)\big],\nonumber\\
\end{eqnarray}
and the matter ($T$-dependent) part reads
\begin{eqnarray}\label{B10}
\lefteqn{\hspace{-1.3cm}
J^{\parallel}_{V,\text{mat}} = \frac{1}{2(2\pi)^{2}}\sum_{\lambda=\pm 1} \lambda  \sum_{\ell=-\infty}^{\infty}\int_{k} \big[ J^{2}_{\ell}(k_{\perp}\rho) - J^{2}_{\ell+1}(k_{\perp}\rho)\big]}\nonumber\\
&&\times \big[n_{\text{f}}(\beta(\omega_{\lambda}+\mu_{\text{eff}})) + n_{\text{f}}(\beta(\omega_{\lambda}-\mu_{\text{eff}})) \big].
\end{eqnarray}
Performing the summation over $\lambda$, $J^{\|}_{V,\text{vac}}$ vanishes and we are left with the matter part of $J^{\parallel}_{V}$. To evaluate the summation over $\ell$ in the matter part, we follow the method first introduced in \cite{ambrus2014,ambrus2019} (see Appendix \ref{appC} for the details of this computation involving a summation over $\ell$ and integrations over $k_{z}$ and $k$). After a lengthy but straightforward computation, we arrive first at
\begin{eqnarray}\label{B11}
\lefteqn{\hspace{-0.2cm}
J^{\parallel}_{V,\text{mat}} = \frac{\Omega T^{2}}{(2\pi)^{2}} \sum_{\lambda = \pm 1}\lambda \sum_{i,r=0}^{\infty} \frac{(2i+2)!}{(2i+2r+1)!}(\rho\Omega)^{2i} (\beta\Omega)^{2r}
}\nonumber\\
&&\times  s^{+}_{r+i,i}\Big[ \text{Li}_{2-2r}(-e^{-\beta(\mu-\lambda\mu_{5})}) - \text{Li}_{2-2r}(-e^{\beta(\mu+\lambda\mu_{5})}) \Big]. \nonumber \\
\end{eqnarray}
Performing the summation over $\lambda$, we arrive at combinations including $\text{Li}_{2-2r}(-e^{x}) +\text{Li}_{2-2r}(-e^{-x})$. As it turns out, this combination is nonvanishing only for $r=0$ and $r=1$. Using \eqref{appB11}, the contribution of $r=1$ vanishes. We are thus left with the contribution from $r=0$. Using
\begin{eqnarray}\label{B12}
\sum_{i=0}^{\infty}(i+1)x^{2i}=\frac{1}{(1-x^{2})^{2}},
\end{eqnarray}
we obtain $J^{\parallel}_{V,\text{mat}}=J^{\parallel}_{V}$ with
\begin{eqnarray}\label{B13}
J^{\parallel}_{V} = \frac{\mu\mu_{5}}{\pi^{2}}\Gamma^{4} \Omega,
\end{eqnarray}
with $\Gamma^{2}\equiv 1/\left( 1-(\rho\Omega)^{2}\right)$. Comparing this result with that appearing, e.g., in \cite{kharzeev2015}, an additional factor $\Gamma^{4}$ appears in \eqref{B13}. This is because in our approach, $J^{\parallel}_{V}$ is computed in the inertial nonrotating frame. As it is known from the Tolman-Ehrenfest theorem \cite{tolman1930-1, tolman1930-2}, in a space described by the metric $g_{\mu\nu}$, the quantities $\mu,\mu_{5}$, $\Omega$, and $T$ transform according to
\begin{eqnarray}\label{B14}
\mu\to \mu'=\mu\Gamma,&\qquad&\mu_{5}\to \mu_{5}^{\prime}=\mu_{5}\Gamma,\nonumber\\
\Omega\to\Omega'=\Omega\Gamma^2,&\qquad&T\to T'=T\Gamma,
\end{eqnarray}
where $\Gamma$ is the Tolman-Ehrenfest factor, defined by $\Gamma\equiv 1/\sqrt{g_{00}}$ with $g_{00}$ the $00$ component of the metric $g_{\mu\nu}$ \cite{ambrus2019-2}. In \eqref{B14}, the quantities $\mu,\mu_{5}$, $\Omega$, and $T$ are quantities in the inertial nonrotating frame and $\mu',\mu'_{5},\Omega'$, and $T'$ are quantities in the noninertial rotating
frame.
\par
Performing a Taylor expansion in the orders of $\Omega$ and considering only the contribution linear in $\Omega$, the chiral vortical coefficient $\sigma_{V}$ defined by $\bs{J}_{V}=\sigma_{V}\bs{\Omega}+\mathcal{O}\left(\Omega^{2}\right)$ is given by
\begin{eqnarray}\label{B15}
\sigma_{V}=\frac{\mu\mu_{5}}{\pi^{2}}.
\end{eqnarray}
Although the same result also appears in \cite{landsteiner2011,landsteiner2013, kharzeev2015}, the methods used in these papers are different from that in the present paper. In what follows, we use the same method to determine the third component  of the axial vector current $J^{\mu}_{A}$ from \eqref{B2}.
\subsection{Axial vector current}\label{sec3B}
Plugging \eqref{A42} into \eqref{B1}, and performing the traces over Dirac $\gamma$ matrices, we arrive after some algebra first at
\begin{eqnarray}\label{B16}
J^{\parallel}_{A} &=&-i\sum_{\lambda=\pm 1}\sum_{\ell=-\infty}^{\infty} \int \frac{dEdk_{z}dk_{\perp}k_{\perp}}{(2\pi)^{3}} \frac{\tilde{E}}{\left(\tilde{E}^{2}-\omega_{\lambda}^{2}\right)}\nonumber\\
&&\times [ J^{2}_{\ell}(k_{\perp}\rho) -  J^{2}_{\ell+1}(k_{\perp}\rho)].
\end{eqnarray}
To introduce finite temperature, we use \eqref{B4} and obtain
\begin{eqnarray}\label{B17}
	J^{\parallel}_{A}&=&\frac{T}{(2\pi)^{2}}\sum_{\lambda=\pm 1}\sum_{\ell=-\infty}^{\infty}\sum_{n=-\infty}^{\infty} \int_{k} \frac{(i\omega_{n} + \mu_{\text{eff}})}{[(i\omega_{n}+\mu_{\text{eff}})^{2}-\omega_{\lambda}^{2}]}\nonumber\\
&&\times [J^{2}_{\ell}(k_{\perp}\rho) -  J^{2}_{\ell+1}(k_{\perp}\rho)].
\end{eqnarray}
Here, the same change of variable $k_{\perp}\to |\bs{k}|$ as in the previous section is performed. To evaluate the summation over $n$, we use the following decomposition:
\begin{eqnarray}\label{B18}
\lefteqn{\sum_{n} \frac{i\omega_{n}}{(i\omega_{n}+\mu_{\text{eff}})^{2} -\omega_{\lambda}^{2} }}\nonumber\\
 &=& \frac12 \sum_{n} \left( \frac1{i\omega_{n}+\mu_{\text{eff}}-\omega_{\lambda} }+ \frac1{i\omega_{n} +\mu_{\text{eff}}+\omega_{\lambda}} \right) \nonumber\\
&&- \sum_{n} \frac{\mu_{\text{eff}}}{(i\omega_{n}+\mu_{\text{eff}})^{2} -\omega_{\lambda}^{2}}.
\end{eqnarray}
Plugging this expression into \eqref{B17} and using
\begin{eqnarray}\label{B19}
\sum_{n=-\infty}^{\infty}\frac{1}{i\omega_{n}-\zeta}\mapsto\beta n_{\text{f}}(\beta\zeta),
\end{eqnarray}
we arrive at
\begin{eqnarray}\label{B20}
J_{A}^{\|}=J_{A,\text{vac}}^{\|}+J_{A,\text{mat}}^{\|},
\end{eqnarray}
where the vacuum ($T$-independent) part is given by
\begin{eqnarray}\label{B21}
J_{A,\text{vac}}^{\|}=\frac{1}{2(2\pi)^{2}}\sum_{\lambda=\pm 1}\sum_{\ell=-\infty}^{\infty}\int_{k}\big[ J^{2}_{\ell}(k_{\perp}\rho) - J^{2}_{\ell+1}(k_{\perp}\rho)\big],\nonumber\\
\end{eqnarray}
and the matter ($T$-dependent) part reads
\begin{eqnarray}\label{B22}
\lefteqn{\hspace{-1.3cm}
J^{\parallel}_{A,\text{mat}} = \frac{-1}{2(2\pi)^{2}}\sum_{\lambda=\pm 1}\sum_{\ell=-\infty}^{\infty}\int_{k} \big[ J^{2}_{\ell}(k_{\perp}\rho) - J^{2}_{\ell+1}(k_{\perp}\rho)\big]}\nonumber\\
&&\times \big[n_{\text{f}}(\beta(\omega_{\lambda}+\mu_{\text{eff}})) - n_{\text{f}}(\beta(\omega_{\lambda}-\mu_{\text{eff}})) \big].
\end{eqnarray}
Let us notice that the summation over $n$ on the lhs of \eqref{B19} diverges. Using the method described in Appendix \ref{appD}, it is possible to determine its finite part, which is given by $\beta n_{\text{f}}(\beta\zeta)$, where $n_{\text{f}}(z)$ is the fermion distribution function.
\par
Using \eqref{appB1}, the vacuum part \eqref{B21} vanishes and we are left with the matter part $J^{\parallel}_{A,\text{mat}}$ from \eqref{B22}. Following the method described in Appendix \ref{appC} and performing the summation over $\ell$ and the integration over $k_{z}$ and $|\bs{k}|$, we arrive after some algebra at
\begin{eqnarray}\label{B23}
\lefteqn{\hspace{-0.2cm}
J^{\parallel}_{A,\text{mat}} = \frac{-\Omega T^{2}}{(2\pi)^{2}} \sum_{\lambda = \pm 1}\sum_{i,r=0}^{\infty} \frac{(2i+2)!}{(2i+2r+1)!}(\rho\Omega)^{2i} (\beta\Omega)^{2r}
}\nonumber\\
&&\times  s^{+}_{r+i,i}\Big[ \text{Li}_{2-2r}(-e^{-\beta(\mu-\lambda\mu_{5})}) + \text{Li}_{2-2r}(-e^{\beta(\mu+\lambda\mu_{5})}) \Big]. \nonumber \\
\end{eqnarray}
Similar to the previous case, the only nonvanishing contributions are from $r=0,1$. Using \eqref{appB11}, \eqref{B12}, and
\begin{eqnarray}\label{B24}
\sum_{i=0}^{\infty}(i+1)(2i+1)x^{2i}=\frac{1+3x^{2}}{(1-x^{2})^{3}},
\end{eqnarray}
we obtain $J^{\parallel}_{A,\text{mat}}=J^{\parallel}_{A}$, with
\begin{eqnarray}\label{B25}
J_{A}^{\parallel}=\left( \frac{T^{2}}{6} + \frac{\mu^{2}+\mu_{5}^{2}}{2\pi^{2}}\right)\Gamma^{4}\Omega + \frac{\Omega^{3}\Gamma^{6}}{24\pi^{2}}\left(1+3(\rho\Omega)^{2}\right). \nonumber\\
\end{eqnarray}
The factors $\Gamma$ appearing on the rhs of \eqref{B25} are consistent with \eqref{B14}, according to the Tolman-Ehrenfest theorem.
The chiral vortical coefficient $\sigma_{A}$ associated with $\bs{J}_{A}=\sigma_{A}\bs{\Omega}+\mathcal{O}\left(\Omega^{2}\right)$ is thus given by
\begin{eqnarray}\label{B26}
\sigma_{A}= \frac{T^{2}}{6} + \frac{\mu^{2}+\mu_{5}^{2}}{2\pi^{2}}.
\end{eqnarray}
The same result also arises in \cite{kharzeev2015}. As is described in Sec. \ref{sec1}, the origin of the $T^{2}$-dependent term in $\sigma_{A}$ is ambiguous. Here, it arises naturally by following the method described in this section. As it is argued in \cite{landsteiner2011,landsteiner2013} it is associated with the gravitational anomaly. The second $(T,\mu,\mu_{5})$ independent term in \eqref{B25} and depending solely on $\Omega$, is unphysical \cite{ambrus2014}. In \cite{ambrus2014}, it is described how these terms arise from the specific Vilenkin (Minkowski) quantization \cite{vilenkin1980}, which is different from Iyer quantization \cite{iyer1982}. In the present paper, we follow Vilenkin's method and introduce the finite temperature by replacing the fermion energy in a corotating system with the Matsubara frequency \cite{kapusta-book}. We notice that the energy in the corotating system $E$  is defined by $E=\tilde{E}-\bs{\Omega}\cdot \bs{J}$ \cite{landau-book}, where $\tilde{E}$ is the energy in the inertial (laboratory) frame and $\bs{J}$ is the total angular momentum, including orbital and spin angular momentum, $\bs{L}$ and $\bs{S}$. At finite temperature, the distinction between these two methods is evident in the alternative substitution of $E$ (Vilenkin quantization) or $\tilde{E}$ (Iyer quantization) with the Matsubara frequencies $i\omega_{n}$.
\section{Moment of inertia of a rigidly rotating Fermi gas}\label{sec4}
\setcounter{equation}{0}
In classical mechanics, the moment of inertia is defined as a response of a classical object
to a finite angular momentum. In this sense, we have
\begin{eqnarray}\label{C1}
\bs{\mathcal{J}}=\mathcal{I}_{\text{tot}}(\Omega,T,\cdots)\bs{\Omega}.
\end{eqnarray}
Here, $\bs{\mathcal{J}}$ is the total angular momentum, consisting of the orbital and spin angular momentum, and $\mathcal{I}_{\text{tot}}$ is the moment of inertia depending in general on $\Omega$, $T$, and other thermodynamical variables. In the first approximation, the moment of inertia $I_{\text{tot}}$ is independent of $\Omega$ and is determined by a Taylor expansion of $\mathcal{I}_{\text{tot}}(\Omega,T,\cdots)$,
\begin{eqnarray}\label{C2}
\mathcal{I}_{\text{tot}}(\Omega,T,\cdots)=I_{\text{tot}}(0,T,\cdots)+\mathcal{O}(\Omega).
\end{eqnarray}
In this section, we determine the total moment of inertia of a rigidly rotating Fermi gas, as well as its moment of inertia corresponding to the orbital and spin angular momenta, $I_{L}$ and $I_{S}$, defined by
\begin{eqnarray}\label{C3}
\bs{j}_{L}=I_{L}\bs{\Omega},\qquad \bs{j}_{S}=I_{S}\bs{\Omega}.
\end{eqnarray}
To define $\bs{j}_{L}$ and $\bs{j}_{S}$, let us consider the total angular momentum density operator \cite{peskin-book, fukushima2018} $\bs{\hat{j}}_{\text{tot}}$, which consists of two parts,
\begin{eqnarray}\label{C4}
\bs{\hat{j}}_{\text{tot}}=\bs{\hat{j}}_{L}+\bs{\hat{j}}_{S},
\end{eqnarray}
the orbital and spin angular momentum operator densities
\begin{eqnarray}\label{C5}
\bs{\hat{j}}_{L}\equiv \bar{\psi}\gamma^{0}\bs{L}\psi,\quad\text{and}\quad \bs{\hat{j}}_{S}\equiv \bar{\psi}\gamma^{0}\bs{S}\psi.
\end{eqnarray}
Here, the orbital angular momentum $\bs{L}$ is given by $\bs{L}=\bs{x}\times \bs{p}$ and the spin operator $\bs{S}$ is defined in \eqref{A14}. Assuming, as in previous sections, that the angular velocity is aligned in the third direction, $\bs{\Omega}=\Omega\bs{e}_{z}$, and having in mind that
\begin{eqnarray}\label{C6}
L_{z}\psi=\left(\ell \mathcal{O}_{+}+(\ell+1)\mathcal{O}_{-}\right)\psi=\left(\ell+\mathcal{O}_{-}\right)\psi,
\end{eqnarray}
where $\psi$ from \eqref{A26} is the solution of Dirac equation \eqref{A9} and $\mathcal{O}_{\pm}$ is defined in Sec. \ref{sec2}, the third component of the orbital angular momentum $\bs{j}_{L}$ is given by
\begin{eqnarray}\label{C7}
\hspace{-0.5cm}j_{L}^{\|}&\equiv& \langle:
\bar{\psi}\gamma^{0}L_{z}\psi:\rangle\nonumber\\
&=&-i\lim\limits_{x\to x^{\prime}}\sum_{\ell=-\infty}^{\infty}\mbox{tr}\left(\gamma^{0}\left(\ell+\mathcal{O}_{-}\right)S_{\ell}(x,x')\right).
\end{eqnarray}
Here, $S_{\ell}$ is given by $S(x,x')=\sum_{\ell}S_{\ell}(x,x')$ with $S(x,x')$ from \eqref{A42}. The subscript $\|$ denotes the third component of the orbital angular momentum density $\bs{j}_{L}$ aligned in the direction of $\bs{\Omega}$ and $:\cdots:$ stands for normal ordering.
\par
As concerns the spin angular momentum density $\bs{j}_{S}$ defined by $\bs{j}_{S}\equiv  \langle:
\bar{\psi}\gamma^{0}\bs{S}\psi:\rangle$, it is possible to use the properties of Dirac $\gamma$ matrices and the definition of $\gamma_{5}$ to show that
$\bs{j}_{S}=\frac{1}{2}\bs{J}_{A}$, where $\bs{J}_{A}$ is the axial vector current \cite{fukushima2018}. For the third component of $\bs{j}_{S}$, we thus have
\begin{eqnarray}\label{C8}
j_{S}^{\|}&\equiv& \langle:
\bar{\psi}\gamma^{0}S_{z}\psi:\rangle=\frac{1}{2}J_{A}^{\|}\nonumber\\
&=&-\frac{i}{2}\lim\limits_{x\to x^{\prime}}\sum_{\ell=-\infty}^{\infty}\mbox{tr}\left(\gamma^{3}\gamma^{5}S_{\ell}(x,x')\right),
\end{eqnarray}
with $J_{A}^{\|}$ defined in \eqref{B2} and given in \eqref{B25}.
Using at this stage, $J_{z}\psi=(\ell+1/2)\psi$ from \eqref{A11}, $\mathcal{O}_{-}=\left(1-i\gamma^{1}\gamma^{2}\right)/2$, and $\gamma^{0}\gamma^{1}\gamma^{2}=\gamma^{3}\gamma^{5}$, we obtain
\begin{eqnarray}\label{C9}
j_{L}^{\|}=j_{\text{tot}}^{\|}-j_{S}^{\|},
\end{eqnarray}
where
\begin{eqnarray}\label{C10}
j_{\text{tot}}^{\|}=-i\lim\limits_{x\to x^{\prime}}\sum_{\ell=-\infty}^{\infty}\mbox{tr}\left(\gamma^{0}(\ell+1/2)S_{\ell}(x,x')\right). \nonumber\\
\end{eqnarray}
In what follows we determine $j_{\text{tot}}^{\|}$ and $j_{S}^{\|}$ using the method in Sec \ref{sec2}. Expanding then the resulting expressions in the orders of $\Omega$, and keeping only the linear term in $\Omega$, yields the total and the spin moment of inertia $I_{\text{tot}}$ and $I_{S}$, defined by
\begin{eqnarray}\label{C11}
j_{\text{tot}}^{\|}=I_{\text{tot}}\Omega, \qquad j_{S}^{\|}=I_{S}\Omega.
\end{eqnarray}
This leads immediately to $I_{L}$ arising from [see \eqref{C4}]
\begin{eqnarray}\label{C12}
I_{L}=I_{\text{tot}}-I_{S}.
\end{eqnarray}
Using \eqref{C8} and \eqref{B25}, the third component of spin angular momentum density reads
\begin{eqnarray}\label{C13}
j_{S}^{\|}&=&\left( \frac{T^{2}}{12} + \frac{\mu^{2}+\mu_{5}^{2}}{4\pi^{2}}\right)\Gamma^{4}\Omega + \frac{\Omega^{3}\Gamma^{6}}{48\pi^{2}}\left(1+3(\rho\Omega)^{2}\right). \nonumber\\
\end{eqnarray}
As it is explained in the previous section, the second term which depends solely on $\Omega$, is unphysical and has its origin in the specific (Vilenkin) quantization that is used in this paper.
Expanding the first term in \eqref{C13} in the orders of $\Omega$ and keeping the $\mathcal{O}(\Omega^0)$ term, the spin part of the moment of inertia $I_{S}$ is given by
\begin{eqnarray}\label{C14}
I_{S}= \frac{T^{2}}{12} + \frac{\mu^{2}+\mu_{5}^{2}}{4\pi^{2}}.
\end{eqnarray}
To determine $j_{L}^{\|}$, we use \eqref{C9}, and we compute $j_{\text{tot}}^{\|}$ from \eqref{C10} by following the same method as described in Sec. \ref{sec3}.
Plugging $S(x,x')$ from \eqref{A42} into \eqref{C10} and performing the traces over Dirac $\gamma$ matrices, we arrive first at
\begin{eqnarray}\label{C15}
j_{\text{tot}}^{\|}&=& -i\sum_{\lambda=\pm 1}\sum_{\ell=-\infty}^{\infty} \int \frac{dEdk_{z}dk_{\perp}k_{\perp}}{(2\pi)^{3}}   \frac{\tilde{E}}{\left(\tilde{E}^{2}-\omega_{\lambda}^{2}\right)}\nonumber\\
&&\times \left(\ell+\frac12\right)[J^{2}_{\ell}(k_{\perp}\rho) + J^{2}_{\ell+1}(k_{\perp}\rho)] .
\end{eqnarray}
Then, introducing the Matsubara frequencies using \eqref{B4} and performing a change of variable $k_{\perp}\to |\bs{k}|$, we obtain
\begin{eqnarray}\label{C16}
j_{\text{tot}}^{\|} &=& T \sum_{\lambda=\pm 1}\sum_{\ell=-\infty}^{\infty}\sum_{n=-\infty}^{\infty} \int_{k}\frac{\left(i\omega_{n}+\mu_{\text{eff}}\right)}{[(i\omega_{n}+\mu_{\text{eff}})^{2}\omega_{\lambda}^{2}]}\nonumber\\
&&\times\left(\ell+\frac12\right) [J^{2}_{\ell}(k_{\perp}\rho) + J^{2}_{\ell+1}(k_{\perp}\rho)].
\end{eqnarray}
Following same steps leading from \eqref{B17} to \eqref{B20}, we get
\begin{eqnarray}\label{C17}
j_{\text{tot}}^{\|}=j_{\text{tot,vac}}^{\|}+j_{\text{tot,mat}}^{\|},
\end{eqnarray}
with the vacuum ($T$-independent) part
\begin{eqnarray}\label{C18}
j^{\|}_{\text{tot,vac}}&=&\frac{1}{2(2\pi)^{2}}\sum_{\lambda=\pm 1}\sum_{\ell=-\infty}^{\infty}\int_{k}\left(\ell+\frac12\right)\nonumber\\
&&\times[J^{2}_{\ell}(k_{\perp}\rho) + J^{2}_{\ell+1}(k_{\perp}\rho)],
\end{eqnarray}
and the matter ($T$-dependent) part
\begin{eqnarray}\label{C19}
\lefteqn{\hspace{-1.3cm}
j^{\|}_{\text{tot,mat}} = -\frac{1}{2(2\pi)^{2}}\sum_{\lambda=\pm 1}\sum_{\ell=-\infty}^{\infty}\int_{k}\left(\ell+\frac12\right)}\nonumber\\
&&\times\big[ J^{2}_{\ell}(k_{\perp}\rho) +J^{2}_{\ell+1}(k_{\perp}\rho)\big]\nonumber\\
&&\times \big[n_{\text{f}}(\beta(\omega_{\lambda}+\mu_{\text{eff}})) - n_{\text{f}}(\beta(\omega_{\lambda}-\mu_{\text{eff}})) \big].
\end{eqnarray}
Using \eqref{appB3},  $j^{\|}_{\text{tot,vac}}$ vanishes. Following at this stage the method described in Appendix \ref{appC} and performing the summation over $\ell$ and the integration over $k_{z}$ and $|\bs{k}|$, we obtain
\begin{eqnarray}\label{C20}
\lefteqn{
	j^{\|}_{\text{tot,mat}} = -\frac{2\Omega T^{2}}{(2\pi)^{2}} \sum_{\lambda=\pm 1}\sum_{i,r=0}^{\infty}\frac1{(2i+1)}\frac{(2i+2)!}{(2r+2i+1)!}}\nonumber\\
&&\times  (\rho\Omega)^{2i} (\beta\Omega)^{2r} s^{+}_{r+i+1,i}\nonumber\\
&&\times \Big[ \text{Li}_{2-2r}(-e^{-\beta(\mu-\lambda\mu_{5})}) +\text{Li}_{2-2r}(-e^{\beta(\mu+\lambda\mu_{5})}) \Big].\nonumber\\
\end{eqnarray}
Similar to previous cases, the only nonvanishing contributions arise from $r=0$ and $r=1$. They are given by
\begin{eqnarray}\label{C21}
j^{\|}_{\text{tot,mat}}\Big|_{r=0} &=& \left(\frac{T^{2}}{36} + \frac{\mu^{2}+\mu_{5}^{2}}{12\pi^{2}}\right)\Gamma^{8}[\left(\rho\Omega\right)^{4}+8(\rho\Omega)^{2}+3]\Omega ,\nonumber\\
j^{\|}_{\text{tot,mat}}\Big|_{r=1} &=& \frac{\Gamma^{10}}{240\pi^{2}}[9(\rho\Omega)^{4}+66(\rho\Omega)^{2}+5]\Omega^{3}.
\end{eqnarray}
To arrive at these results, we have used
\begin{eqnarray}\label{C22}
\sum_{i=0}^{\infty}\frac{(i+1)}{2i+1}s_{i+1,i}^{+}x^{2i}&=&\frac{x^{4}+8x^{2}+3}{12(1-x^{2})^{4}},\nonumber\\
\sum_{i=0}^{\infty}\frac{1}{(2i+1)(2i+3)}s_{i+2,i}^{+}x^{2i}&=&\frac{9x^{4}+66x^{2}+5}{240(1-x^{2})^{5}}, \nonumber\\
\end{eqnarray}
with $s^{+}_{i+1,i}$ and $s_{i+2,i}^{+}$ from \eqref{appB7}. Omitting unphysical terms arising from the $r=1$ contribution in \eqref{C21}, expanding remaining terms in the orders of $\Omega$, and considering only the coefficient of the term linear in $\Omega$, the total angular momentum density of a rigidly rotating free Fermi gas reads
\begin{eqnarray}\label{C23}
j_{\text{tot}}^{\|}=\left(\frac{T^{2}}{12} + \frac{\mu^2+\mu_5^2}{4\pi^{2}}\right)\Omega+\mathcal{O}(\Omega^{3}).
\end{eqnarray}
The total moment of inertia of this gas is thus given by
\begin{eqnarray}\label{C24}
I_{\text{tot}}=\frac{T^{2}}{12} + \frac{\mu^{2}+\mu_{5}^{2}}{4\pi^{2}}.
\end{eqnarray}
Plugging \eqref{C24} and \eqref{C14} into \eqref{C12}, the moment of inertia corresponding to orbital angular momentum density vanishes,
\begin{eqnarray}\label{C25}
I_{L}=0.
\end{eqnarray}
Thus, the total moment of inertia is solely given by $I_{S}$. It would be intriguing to find a circumstance under which $I_{L}$ of a rigidly rotating Fermi gas does not vanish.
\par
Let us notice that the covariant total angular momentum $J_{C}^{\lambda,\mu\nu}$ is defined by
\begin{eqnarray}\label{C26}
J_{C}^{\lambda,\mu\nu}=x^{\mu}T_{C}^{\lambda\nu}-x^{\nu}T_{C}^{\lambda\mu}+S_{C}^{\lambda,\mu\nu},
\end{eqnarray}
where $T_{C}^{\mu\nu}$ is the canonical energy-momentum tensor and $S_{C}^{\lambda,\mu\nu}$ is the canonical spin given by
\begin{eqnarray}\label{C27}
S_{C}^{\lambda,\mu\nu}=-\frac{1}{2}\epsilon^{\lambda\mu\nu\alpha}\bar{\psi}\gamma_{\alpha}\gamma_{5}\psi.
\end{eqnarray}
In Appendix \ref{appD2}, we show that $S_{C}^{i}\equiv \frac{1}{2}\epsilon^{ijk}S_{C}^{0,jk}$ is equal to $\langle:\hat{j}_{S}^{i}:\rangle$, where $\bs{\hat{j}}_{S}$ is defined in \eqref{C5}. As we have stated in Sec. \ref{sec1}, the decomposition of total angular momentum into orbital and spin parts is affected by ambiguities arising from the possibility of redefining the spin part of the total angular momentum through pseudogauge transformation \cite{speranza2021}.
Here, it is argued that HW and GLW pseudogauges are only relevant for massive fermions. The canonical and BR pseudogauges, which are applicable for massless fermions, yield different results. In this section, we used the canonical pseudogauge for massless fermions and showed that $I_{\text{tot}}=I_{S}$ and $I_{L}=0$. 
\section{Concluding remarks}\label{sec5}
\setcounter{equation}{0}
In this paper, we studied the effect of rotation on vector and axial vector currents in a rigidly rotating and chirally imbalanced free Fermi gas. We first solved the Dirac equation of rigidly rotating fermions at finite chemical potential $\mu$ and axial chemical potential $\mu_{5}$. Because of the symmetry properties of the Dirac equation, the solutions consist of plane waves in three directions and a Bessel function in the radial direction. We then used the standard Fock-Schwinger method and determined the free fermion propagator of this model. Using the fact that the vector and axial vector currents can be expressed in terms of free fermion propagator, it was possible to determine these currents in terms of temperature $T$, $\mu$, $\mu_{5}$, and the angular velocity $\Omega$.
Expanding the resulting expressions in the orders of $\Omega$, we determined the chiral vortical conductivities $\sigma_{V}$ and $\sigma_{A}$, associated with the vector and axial vector currents. The resulting expressions coincided with $\sigma_{V}$ and $\sigma_{A}$, which were determined using various other methods (see \cite{kharzeev2015} and the references therein). We showed that $\sigma_{V}$ is proportional to $\mu$ and $\mu_{5}$, and $\sigma_{A}$ includes terms proportional to $T^{2}, \mu^{2},$ and $\mu_{5}^{2}$.
The final result for $J_{A}^{\|}$ included, apart from physical parts an unphysical part, which arose, according to the arguments presented in \cite{ambrus2014}, from the difference between the quantization of rotating fermions by Vilenkin \cite{vilenkin1980} and Iyer \cite{iyer1982}, as discussed in Sec. \ref{sec3B}.
\par
Finally, motivated by recent results from \cite{chernodub2023} concerning the possibility for a realization of negative Barnett effect in rotating fermionic gases, we focused on the moment of inertia of a rigidly rotating free Fermi gas and presented a method for computing the moments of inertia associated with the orbital angular momentum, $I_{L}$, and the spin angular momentum, $I_{S}$, separately.
First, we used a relation between the spin angular momentum density and the axial vector current  to derive $I_{S}$ directly [see \eqref{C8}]. Next, we calculated the total angular momentum density, which can similarly be expressed in terms of the free fermion propagator [see \eqref{C10}]. By applying the method introduced in Sec. \ref{sec3}, we determined the total moment of inertia, $I_{\text{tot}}$. Using the relationship $I_{L} = I_{\text{tot}} - I_{S}$, we subsequently derived $I_{L}$. We demonstrated that $I_{S} = I_{\text{tot}}$ implying that $I_{L}$ vanished for massless fermions. This result is, however, affected by the choice of various pseudogauges. In this paper, we worked with a canonical pseudogauge and arrived at  $I_{\text{tot}} = I_{S}$ and $I_{L}=0$, as aforementioned. 
\par
The fermion propagators derived in the first part of this paper can be used to determine the thermodynamic potential of a rigidly rotating Fermi gas with chiral imbalance. Using this potential, various thermodynamic properties of this gas can be studied. Moreover, investigating the effects of uniform background magnetic fields on a rotating Fermi gas, particularly in relation to chiral vortical conductivities, would be interesting. One potential aim of this research is to explore the interplay among various parameters such as temperature $T$, chemical potential $\mu$, axial chemical potential $\mu_{5}$, angular velocity $\Omega$, and the magnetic field strength $B$. Additionally, an important question is to identify specific conditions under which $I_{\text{tot}} \neq I_{S}$ and $I_{L}$ does not equal zero. Returning to the findings in \cite{chernodub2023}, it would be interesting to examine the signs of $I_{S}$ and $I_{L}$ for a potential realization of the negative Barnett effect in fermionic systems.
\begin{appendix}
\section{Orthonormality relation in cylinder coordinate system}\label{appA}
\setcounter{equation}{0}
In Sec. \ref{sec2}, we presented the fermion propagator in momentum space. We used a Bessel-Fourier transformation \eqref{A43}. The orthonormality relations that were used to derive $S_{\ell\ell'}(p,p')$ from \eqref{A45} and \eqref{A46} are given by
\begin{eqnarray}\label{appA1}
\int dt~e^{-i\left(E-E^{\prime}\right)t}&=&2\pi\delta\left(E-E^{\prime}\right), \nonumber\\
\int dz~e^{i\left(p_{z}-p^{\prime}_{z}\right)z}&=&2\pi\delta\left(p_{z}-p^{\prime}_{z}\right),\nonumber\\
\int_{0}^{2\pi}d\varphi e^{i\left(\ell-\ell'\right)\varphi}&=&2\pi\delta_{\ell,\ell'},\nonumber\\
\int_{0}^{\infty }dr~rJ_{\ell }\left(p_{\perp }r \right)J_{\ell }\left(p'_{\perp }r \right) &=& \frac{1}{p_{\perp }}\delta \left(p_{\perp }-p'_{\perp } \right).
\end{eqnarray}
\section{Useful formula}\label{appB}
\setcounter{equation}{0}
\subsection{Summation over $\bs{\ell}$}\label{appB1}
To evaluate the summation over $\ell$ in the expressions arising in Secs. \ref{sec3} and \ref{sec4}, we use the following identities, which were originally introduced and derived in  \cite{ambrus2014,ambrus2019}.
The first useful relation is given by
\begin{eqnarray}\label{appB1}
\sum_{\ell=-\infty}^{\infty}J_{\ell}^{2}(z)=1.
\end{eqnarray}
Introducing
\begin{eqnarray}\label{appB2}
J_{\ell}^{\pm}(z)\equiv J_{\ell}^{2}(z)\pm J_{\ell+1}^{2}(z),
\end{eqnarray}
it is possible to show
\begin{eqnarray}\label{appB3}
\sum_{\ell=-\infty}^{\infty} \left(\ell+\frac{1}{2}\right)^{2n}J^{+}_{\ell}(z) &=& \sum_{i=0}^{n} \frac{2\Gamma(i+1/2)}{i!\sqrt{\pi}} s^{+}_{n,i} z^{2i},
\nonumber\\
\sum_{\ell=-\infty}^{\infty} \left(\ell+\frac{1}{2}\right)^{2n+1}J^{+}_{\ell}(z) &=&0,
\end{eqnarray}
as well as
\begin{eqnarray}\label{appB4}
\sum_{\ell=-\infty}^{\infty}(\ell+\frac{1}{2})^{2n} J^{-}_{\ell}(z)&=&0,\nonumber\\
\sum_{\ell=-\infty}^{\infty}\left(\ell+\frac{1}{2}\right)^{2n+1} J^{-}_{\ell}(z) &=& \sum_{i=0}^{n} \frac{2\Gamma(i+1/2)}{i!\sqrt{\pi}} s^{-}_{n,i} z^{2i} ,\nonumber\\
\end{eqnarray}
where $s^{\pm}_{n,i}$ with $i<n$, defined by
\begin{eqnarray}\label{appB5}
s^{+}_{n,i} &=& 2\sum_{\ell=0}^{i} \frac{(-1)^{i-\ell}}{(i-\ell)!(i+\ell+1)!} \left(\ell+\frac{1}{2}\right)^{2n+1}  \nonumber \\
&=& \frac{1}{\left(2i+1\right)!}\lim_{\alpha\rightarrow 0} \frac{d^{2n+1}}{d\alpha^{2n+1}}\left(2\sinh\left(\frac{\alpha}{2}\right)\right)^{2i+1},\nonumber \\
\end{eqnarray}
and
\begin{equation}\label{appB6}
s^{-}_{n,i} = \left(i+\frac12\right)s^{+}_{n,i}.
\end{equation}
The latter arises from $\frac{d}{dz} \Big( zJ^{+}_{\ell}(z) \Big) = (2\ell+1) J^{-}_{\ell}(z)$.
In \eqref{appB5}, $s^{+}_{n,i}=0$ for $i>n$. In Secs. \ref{sec3} and \ref{sec4}, we particularly use
\begin{eqnarray}\label{appB7}
s^{+}_{i,i} &=& 1,\nonumber\\
s^{+}_{i+1,i}&=&\frac{1}{24} (2i+1)(2i+2)(2i+3),\nonumber\\
s^{+}_{i+2,i}&=&\frac{1}{5760} (2i+1)(2i+2)(2i+3)(2i+4)(2i+5)\nonumber\\
&&\times(10i+3).
\end{eqnarray}
\subsection{Integrals over Fermi-Dirac distribution function}\label{appB2}
In Sec. \ref{sec3}, we encounter integrals over Fermi-Dirac distribution functions in the form,
\begin{equation}\label{appB8}
I_{2n,r}^{\pm}= \int_{0}^{\infty} d\omega \omega^{r} \frac{d^{2n}}{d\omega^{2n} }n_{\text{f}}\left(\beta(\omega\pm \mu)\right) .
\end{equation}
To evaluate them, we perform a change of variable $u\equiv\beta\omega$ and use the symmetry properties of the distribution function with respect to $\omega$ and $\mu$ to replace the derivation with respect to $u$ with the derivation with respect to $\mu$. We thus have,
\begin{eqnarray}\label{appB9}
I_{2n,r}^{\pm} &=& \beta^{-r-1} (-1)^{2n}\frac{d^{2n}}{d\mu^{2n}} \int_{0}^{\infty } du
\frac{u^{r}}{e^{u\pm\beta\mu}+1}  \nonumber \\
&=& -r! \beta^{2n-r-1} \text{Li}_{r+1-2n}(-e^{\mp\beta\mu}),
\end{eqnarray}
where $\text{Li}_{s}(z)\equiv \sum_{k=1}^{\infty}z^{k}/k^{s}$ is the polylogarithm function of order $s$. Following the same steps, we also have
\begin{equation}\label{appB10}
I_{2n+1,r}^{\pm} = r! \beta^{2n-r} \text{Li}_{r-2n}(-e^{\mp \beta\mu}).
\end{equation}
In Secs. \ref{sec3} and \ref{sec4}, we particularly use \cite{ambrus2019}
\begin{eqnarray}\label{appB11}
\text{Li}_{0}(-e^{+a})+\text{Li}_{0}(-e^{-a}) &=& -1,\nonumber\\
\text{Li}_{2}(-e^{+a})+\text{Li}_{2}(-e^{-a}) &=&-\frac{\pi^{2}}{6} - \frac{a^{2}}{2}.
\end{eqnarray}
\begin{widetext}
\section{Derivation of \eqref{B11}, \eqref{B23}, and \eqref{C20}}\label{appC}
\setcounter{equation}{0}
In Secs. \ref{sec3} and $\ref{sec4}$, we arrive at $J_{V}^{\|}=J_{V,\text{mat}}^{\|}$, $J_{A}^{\|}=J_{A,\text{mat}}^{\|}$, and $j_{\text{tot,mat}}^{\|}$ from \eqref{B11}, \eqref{B23}, and \eqref{C19} with
\begin{eqnarray}
J^{\parallel}_{V,\text{mat}}&=&\frac{1}{2(2\pi)^{2}}\sum_{\lambda} \lambda  \sum_{\ell=-\infty}^{\infty}\int_{k} \big[ J^{2}_{\ell}(k_{\perp}\rho) - J^{2}_{\ell+1}(k_{\perp}\rho)\big]\big[n_{\text{f}}(\beta(\omega_{\lambda}+\mu_{\text{eff}})) + n_{\text{f}}(\beta(\omega_{\lambda}-\mu_{\text{eff}})) \big],\label{appC1}\\
J^{\parallel}_{A,\text{mat}}&=& \frac{-1}{2(2\pi)^{2}}\sum_{\lambda}\sum_{\ell=-\infty}^{\infty}\int_{k} \big[ J^{2}_{\ell}(k_{\perp}\rho) - J^{2}_{\ell+1}(k_{\perp}\rho)\big]\big[n_{\text{f}}(\beta(\omega_{\lambda}+\mu_{\text{eff}})) - n_{\text{f}}(\beta(\omega_{\lambda}-\mu_{\text{eff}})) \big],\label{appC2}\\
j^{\parallel}_{\text{tot,mat}}&=& \frac{-1}{2(2\pi)^{2}}\sum_{\lambda}\sum_{\ell=-\infty}^{\infty}\int_{k} j \big[ J^{2}_{\ell}(k_{\perp}\rho) +J^{2}_{\ell+1}(k_{\perp}\rho)\big] \big[n_{\text{f}}(\beta(\omega_{\lambda}+\mu_{\text{eff}})) - n_{\text{f}}(\beta(\omega_{\lambda}-\mu_{\text{eff}})) \big],\label{appC3}
\end{eqnarray}
from \eqref{B10} and \eqref{B22}. To evaluate these expressions, we expand the fermionic particle distribution function in the order of $\Omega$,
\begin{eqnarray}\label{appC4}
n_{\text{f}}\left(\beta(\omega_{\lambda}\pm\mu_{\text{eff}})\right) = \sum_{r=0}^{\infty} \frac{(\pm\Omega)^{r}}{r!} j^{r} \frac{d^{r}}{d\omega_{\lambda}^{r} } n_{\text{f}}\left(\beta\left(\omega_{\lambda}\pm \mu\right)\right),\nonumber\\
\end{eqnarray}
with $j=\ell+1/2$. Plugging then
\begin{eqnarray}\label{appC5}
n_{\text{f}}(\beta(\omega_{\lambda}+\mu_{\text{eff}})) \pm n_{\text{f}}(\beta(\omega_{\lambda}-\mu_{\text{eff}}))&=&\sum_{r=0}^{\infty}\frac{\Omega^{2r}}{(2r)!}j^{2r}\frac{d^{2r}}{d\omega_{\lambda}^{2r}}\big[n_{\text{f}}(\beta(\omega_{\lambda}+\mu))\pm n_{\text{f}}(\beta(\omega_{\lambda}-\mu))\big]\nonumber\\
&&
+\sum_{r=0}^{\infty}\frac{\Omega^{2r+1}}{(2r+1)!}j^{2r+1}\frac{d^{2r+1}}{d\omega_{\lambda}^{2r+1}}\big[n_{\text{f}}(\beta(\omega_{\lambda}+\mu))\mp n_{\text{f}}(\beta(\omega_{\lambda}-\mu))\big],
\end{eqnarray}
into \eqref{appC1} and \eqref{appC2} and using \eqref{appB4} as well as \eqref{appB3} to perform the summation over $\ell$, we arrive at
\begin{eqnarray}
J^{\parallel}_{V,\text{mat}} &=&+\sum_{\lambda}\lambda\sum_{r=0}^{\infty}\sum_{i=0}^{r} \int_{k}\frac{\Omega^{2r+1}}{(2r+1)!}\frac{\Gamma(i+1/2)}{(2\pi)^{2}i!\sqrt{\pi}}s_{r,i}^{-}(k_{\perp}\rho)^{2i}\frac{d^{2r+1}}{d\omega_{\lambda}^{2r+1}}\big[n_{\text{f}}(\beta(\omega_{\lambda}+\mu))-n_{\text{f}}(\beta(\omega_{\lambda}-\mu))\big],\label{appC6}\\
J^{\parallel}_{A,\text{mat}} &=&-\sum_{\lambda}\sum_{r=0}^{\infty}\sum_{i=0}^{r} \int_{k}\frac{\Omega^{2r+1}}{(2r+1)!}\frac{\Gamma(i+1/2)}{(2\pi)^{2}i!\sqrt{\pi}}s_{r,i}^{-}(k_{\perp}\rho)^{2i}\frac{d^{2r+1}}{d\omega_{\lambda}^{2r+1}}\big[n_{\text{f}}(\beta(\omega_{\lambda}+\mu))+n_{\text{f}}(\beta(\omega_{\lambda}-\mu))\big],\label{appC7}\\
j^{\parallel}_{\text{tot,mat}} &=&-\sum_{\lambda}\sum_{r=0}^{\infty}\sum_{i=0}^{r+1} \int_{k}\frac{\Omega^{2r+1}}{(2r+1)!}\frac{\Gamma(i+1/2)}{(2\pi)^{2}i!\sqrt{\pi}}s_{r+1,i}^{+}(k_{\perp}\rho)^{2i}\frac{d^{2r+1}}{d\omega_{\lambda}^{2r+1}}\big[n_{\text{f}}(\beta(\omega_{\lambda}+\mu))+n_{\text{f}}(\beta(\omega_{\lambda}-\mu))\big],\label{appC8}
\end{eqnarray}
with $s_{r,i}^{-}$ defined in \eqref{appB6}. A change of variables $r\to r+i$ leads to
\begin{eqnarray}
J^{\parallel}_{V,\text{mat}} &=&+\sum_{\lambda}\lambda \sum_{i,r=0}^{\infty}\int_{k}\frac{\Omega^{2r+2i+1}}{(2r+2i+1)!}\frac{\Gamma\left(i+1/2\right)}{(2\pi)^{2}i!\sqrt{\pi}}\left(i+\frac12\right)s_{r+i,i}^{+}(k_{\perp}\rho)^{2i}\frac{d^{2r+2i+1}}{d\omega_{\lambda}^{2r+2i+1}}\big[n_{\text{f}}(\beta(\omega_{\lambda}+\mu))-n_{\text{f}}(\beta(\omega_{\lambda}-\mu))\big],\nonumber\\ \label{appC9}\\
J^{\parallel}_{A,\text{mat}} &=&-\sum_{\lambda}\sum_{i,r=0}^{\infty}\int_{k}\frac{\Omega^{2r+2i+1}}{(2r+2i+1)!}\frac{\Gamma\left(i+1/2\right)}{(2\pi)^{2}i!\sqrt{\pi}}\left(i+\frac12\right)s_{r+i,i}^{+}(k_{\perp}\rho)^{2i}\frac{d^{2r+2i+1}}{d\omega_{\lambda}^{2r+2i+1}}\big[n_{\text{f}}(\beta(\omega_{\lambda}+\mu))+n_{\text{f}}(\beta(\omega_{\lambda}-\mu))\big],\nonumber\\ \label{appC10}\\
j^{\parallel}_{\text{tot,mat}} &=&-\sum_{\lambda}\sum_{i,r=0}^{\infty}\int_{k}\frac{\Omega^{2r+2i+1}}{(2r+2i+1)!}\frac{\Gamma\left(i+1/2\right)}{(2\pi)^{2}i!\sqrt{\pi}}\frac{1}{(i+1/2)}s_{r+i+1,i}^{+}(k_{\perp}\rho)^{2i}\frac{d^{2r+2i+1}}{d\omega_{\lambda}^{2r+2i+1}}\big[n_{\text{f}}(\beta(\omega_{\lambda}+\mu))+n_{\text{f}}(\beta(\omega_{\lambda}-\mu))\big],\nonumber\\\label{appC11}
\end{eqnarray}
with $\int_{k}$ from \eqref{B6}. We perform, at this stage, the integration over $k_{z}$ by utilizing
\begin{eqnarray}\label{appC12}
\int_{-|\bs{k}|}^{+|\bs{k}|}dk_{z} k_{\perp}^{2i}=\frac{\sqrt{\pi}\Gamma(i+1)}{\Gamma(i+3/2)}|\bs{k}|^{2i+1}.
\end{eqnarray}
Plugging this expression into \eqref{appC11}, we are left with the integration over $|\bs{k}|$. To perform this integration, we use \eqref{appB10} and arrive at \eqref{B11}, \eqref{B23}, and \eqref{C20}. Let us notice that in \eqref{appC12} $k_{\perp}^{2}=|\bs{k}|^{2}-k_{z}^{2}$ is used.
\end{widetext}
\section{Technical comments}\label{appD}
\subsection{Derivation of \eqref{B19}}\label{appD1}
\setcounter{equation}{0}
In this section, we will evaluate the sum over $n$ in
\begin{eqnarray}\label{appD1}
\mathcal{S}=\sum_{n=-\infty}^{+\infty}\frac{1}{i\omega_{n}-\zeta},
\end{eqnarray}
where $\omega_{n}=(2n+1)\pi T$. To perform \eqref{appD1}, we follow the standard method from \cite{kapusta-book}, where the summation over $n$ in \eqref{appD1} is regarded as a summation over the residues of a function, which has poles at fermionic Matsubara frequencies $\omega_{n}$. To perform it, we first replace the above summation with a complex integral over a closed contour $C_1$ including the Matsubara frequencies lying on the imaginary axis
\begin{eqnarray}\label{appD2}
\mathcal{S}=\frac{1}{2\pi i}\oint_{C_1} g(z)h(z) dz,
\end{eqnarray}
with $g(z)=(z-\zeta)^{-1}$, and $h(z)$ is a weighting function having simple poles at $z=i\omega_{n}$.
At this stage, it is possible to give $\mathcal{S}$ by an integration over the whole complex plane $C$, including poles of both $g(z)$ and $h(z)$ minus an integration over a closed contour $C_{2}$, including only the pole of $g(z)$:
\begin{eqnarray}\label{appD3}
\mathcal{S}=\frac{1}{2\pi i}\oint_{C}g(z)h(z) dz-\frac{1}{2\pi i}\oint_{C_{2}}g(z)h(z) dz.
\end{eqnarray}
The first integral over $C$ diverges and the second integral is finite. Considering only the finite term and choosing the Fermi-Dirac distribution function $h(z)=-\beta n_{f}(\beta z)$, we arrive at
\begin{eqnarray}\label{appD4}
\mathcal{S}\mapsto\beta n_{\text{f}}(\beta \zeta),
\end{eqnarray}
as stated in \eqref{B19}.
\subsection{The relation between canonical spin tensor $\bs{S^{\mu\nu}}$ and spin angular momentum density $\bs{\hat{j}_{S}}$}\label{appD2}
\setcounter{equation}{0}
According to \cite{speranza2021}, the canonical spin tensor is defined by
\begin{eqnarray}\label{appD5}
S_{C}^{\lambda,\mu\nu}=\frac{1}{4}\bar{\psi}\{\gamma^{\lambda},\sigma^{\mu\nu}\}\psi=-\frac{1}{2}\epsilon^{\lambda\mu\nu\alpha}\bar{\psi}\gamma_{\alpha}\gamma_{5}\psi.
\end{eqnarray}
Let us consider $S_{C}^{0,ij}$. It is given by
\begin{eqnarray}\label{appD6}
S_{C}^{0,ij}=-\frac{1}{2}\epsilon^{ijk}\bar{\psi}\gamma_{k}\gamma_{5}\psi,
\end{eqnarray}
where $\epsilon^{0ijk}=\epsilon^{ijk}$ is used. Multiplying this expression with $\frac{1}{2}\epsilon^{ijm}$ and using $\epsilon^{ijm}\epsilon^{ijk}=-2\delta^{mk}$, we arrive at
\begin{eqnarray}\label{appD7}
S_{C}^{m}\equiv \frac{1}{2}\epsilon^{ijm}S_{C}^{0,ij}=\frac{1}{2}\bar{\psi}\gamma^{m}\gamma_{5}\psi=\frac{1}{2}J_{A}^{m}, \qquad m=1,2,3.\nonumber\\
\end{eqnarray}
In Sec. \ref{sec4}, we denote $\bs{S}_{C}$ by $\langle:\bs{\hat{j}}_{S}:\rangle$ with spin angular momentum density operator $\bs{\hat{j}}_{S}=\bar{\psi}\gamma^{0}\bs{S}\psi$  from \eqref{C5}. Using the definition of $\bs{S}=\frac{1}{2}\gamma^{5}\gamma^{0}\bs{\gamma}$, it is possible to show that $\langle :\bs{j}_{S}:\rangle=\frac{1}{2}\bs{J}_{A}$, as stated in \eqref{C8}.
\end{appendix}

\end{document}